# Anchor-Controlled Generative Adversarial Network for High-Fidelity Electromagnetic and Structurally Diverse Metasurface Design

Yunhui Zeng[1,2], Hongkun Cao[2*], Xin Jin[1,2*],

[1]Shenzhen International Graduate School, Tsinghua University, Shenzhen 518055, China

[2]Peng Cheng Laboratory, Shenzhen 518055, China

**\*Email:** jin.xin@sz.tsinghua.edu.cn (Xin Jin), caohk@pcl.ac.cn (Hongkun Cao)

**ABSTRACT**: Metasurfaces, capable of manipulating light at subwavelength scales, hold great potential for advancing optoelectronic applications. Generative models, particularly Generative Adversarial Networks (GANs), offer a promising approach for metasurface inverse design by efficiently navigating complex design spaces and capturing underlying data patterns. However, existing generative models struggle to achieve high electromagnetic fidelity and structural diversity. These challenges arise from the lack of explicit electromagnetic constraints during training, which hinders accurate structure-to-electromagnetic response mapping, and the absence of mechanisms to handle one-to-many mappings dilemma, resulting in insufficient structural diversity. To address these issues, we propose the Anchor-controlled Generative Adversarial

Network (AcGAN), a novel framework that improves both electromagnetic fidelity and structural diversity. To achieve high electromagnetic fidelity, AcGAN proposes the Spectral Overlap Coefficient (SOC) for precise spectral fidelity assessment and develops AnchorNet, which provides real-time feedback on electromagnetic performance to refine the structure-to-electromagnetic mapping. To enhance structural diversity, AcGAN incorporates a cluster-guided controller that refines input processing and ensures multi-level spectral integration, guiding the generation process to explore multiple configurations for the same spectral target. Additionally, a dynamic loss function progressively shifts the focus from data-driven learning to optimizing both spectral fidelity and structural diversity. Empirical analysis shows that AcGAN reduces the Mean Squared Error (MSE) by 73% compared to current state-of-the-art GANs methods and significantly expands the design space to generate diverse metasurface architectures that meet precise spectral demands. This breakthrough demonstrates AcGAN's transformative impact on metasurface design, providing a robust framework for advanced optoelectronic applications.

Metasurfaces constructed of two-dimensional artificial material structures at subwavelength scales have garnered significant attention for their unparalleled ability to manipulate intrinsic properties of light, including spectrum[1,2], amplitude[3,4], phase[5,6], polarization[7,8], and wavefront[9,10]. This extraordinary capability arises from the vast design flexibility afforded by the spatial and material configurations of meta-atoms, enabling functionalities far beyond those of natural materials. Leveraging this design flexibility, recent advancements in metasurface design have led to the realization of novel functionalities such as light field imaging[11], holographic display[12], perfect absorption[13], vortex beam generation[14], optical encryption[15], and optical communication[16], showcasing the potential of metasurfaces to revolutionize optical technologies.

Even though modern numerical methods allow for the calculation of the electromagnetic (EM) response of complex structures and diverse materials, the design of metasurfaces is still challenging owing to the nonintuitive and nonunique relationship between physical structures, material properties, and their EM responses[17]. Traditionally, metasurface design relies on physics-inspired methods and human expertise, including insights from analytical models, experience from previous designs, and scientific intuition. Techniques such as resonant phase control[18], propagation phase control[19], and geometric phase control[20], used independently or collectively[21,22], are pivotal for precise phase response tailoring. However, these methods constrain the design space, limiting innovation primarily to meta-atom configurations, which highlights a related shortcoming: the fundamental theory underpinning metasurfaces is not yet well-established[23]. As design complexity increases, the traditional expert-knowledge-based paradigm becomes less effective[24]. Furthermore, the widely used trial-and-error method, combined with extensive scanning, is constrained by its limited optimization space and the time-consuming process of solving Maxwell's equations[25].

Deep learning (DL), a subset of artificial intelligence (AI), has emerged as a transformative tool for metasurface design, effectively addressing the challenges posed by traditional methods. By mapping the complex relationships between metasurface parameters and their EM responses, DL facilitates direct design processes while significantly reducing reliance on computationally expensive simulations[26]. Among various DL-driven approaches, Generative Adversarial Networks (GANs) [27,28] stand out due to their capacity to learn intricate data distributions and generate diverse metasurface structures. This capability not only alleviates the limitations inherent in expert-knowledge-based paradigms—such as their restricted design space and dependence on trial-and-error methodologies—but also paves the way for enhanced design flexibility and innovation in metasurface design. However, GANs often generate outputs that resemble training data without precise control over specific characteristics. To address this, Conditional Generative Adversarial Networks (CGANs)[29] address this limitation by introducing conditional inputs, enabling the generation of designs that can align more closely with predefined EM characteristics[30]. Nonetheless, GAN-based methods still have two critical challenges to address: i) **Limited Electromagnetic Fidelity**: GAN-based methods typically focus on generating visually accurate structures, but often lack explicit constraints to ensure high EM fidelity. This deficiency stems from the absence of direct feedback on EM performance during training, making it difficult for models to learn the complex mapping between metasurfaces and their EM responses. As a result, generated designs may align with the visual characteristics of the dataset but fail to meet the precise EM requirements. ii) **Limited Structural Diversity**: Metasurface design involves a one-to-many mapping dilemma, where multiple structures can produce the same EM response. However, GAN-based methods often generate limited structural diversity, predominantly producing solutions that resemble the most frequently observed configurations in the training data. This limitation arises from the lack of mechanisms that facilitate the exploration of diverse configurations capable of achieving the same EM targets, thus limiting the potential diversity of the generated metasurface designs. The resulting lack of structural diversity critically impairs the adaptability of designs to a range of application requirements and manufacturing constraints. This deficiency may obstruct the identification of optimal structures that could enhance performance or address specific functional demands, ultimately undermining the robustness and applicability of metasurfaces in practical implementations.

In this study, we focus on the complex task of designing free-form metasurface filters using AI to control and enhance spectral absorption, as demonstrated in **Figure 1**. To navigate the complex inverse design problem that balances both material and structural properties, we utilize an encoding strategy where key metasurface parameters—including refractive indices, plasma frequencies, and resonator geometries—are mapped into discrete "RGB" channels of color images, capturing a broad design space. To achieve high EM fidelity, our proposed Anchor-controlled Generative Adversarial Network (AcGAN) proposes the Spectral Overlap Coefficient (SOC), a novel metric developed to evaluate the alignment between the generated and target spectral responses, thereby ensuring precise control over the spectral characteristics of the metasurfaces. Furthermore, we develop AnchorNet, a predictive model embedded in the generative framework, provides real-time feedback on EM performance during training. This feedback mechanism significantly improves the model's ability to optimize the complex structure-to-EM mapping. For enhancing structural diversity, AcGAN proposes a cluster-guided controller that promotes the exploration of multiple valid configurations for any given spectral target, effectively addressing the one-to-many mapping dilemma inherent in metasurface design. Combined with our dynamic loss function, this approach shifts the focus from initial data-driven learning to a more balanced optimization of both spectral fidelity and structural diversity. These collective advancements empower AcGAN to not only bridge the gap between visual resemblance and functional EM performance but also establish a robust framework for designing metasurfaces that meet stringent requirements for high EM fidelity and structural diversity in advanced optoelectronic applications. Empirical analysis demonstrates that AcGAN significantly reduces the Mean Squared Error (MSE) by 73% compared to current state-of-the-art GAN methods and markedly expands the design space to generate diverse metasurface architectures that meet precise spectral demands.

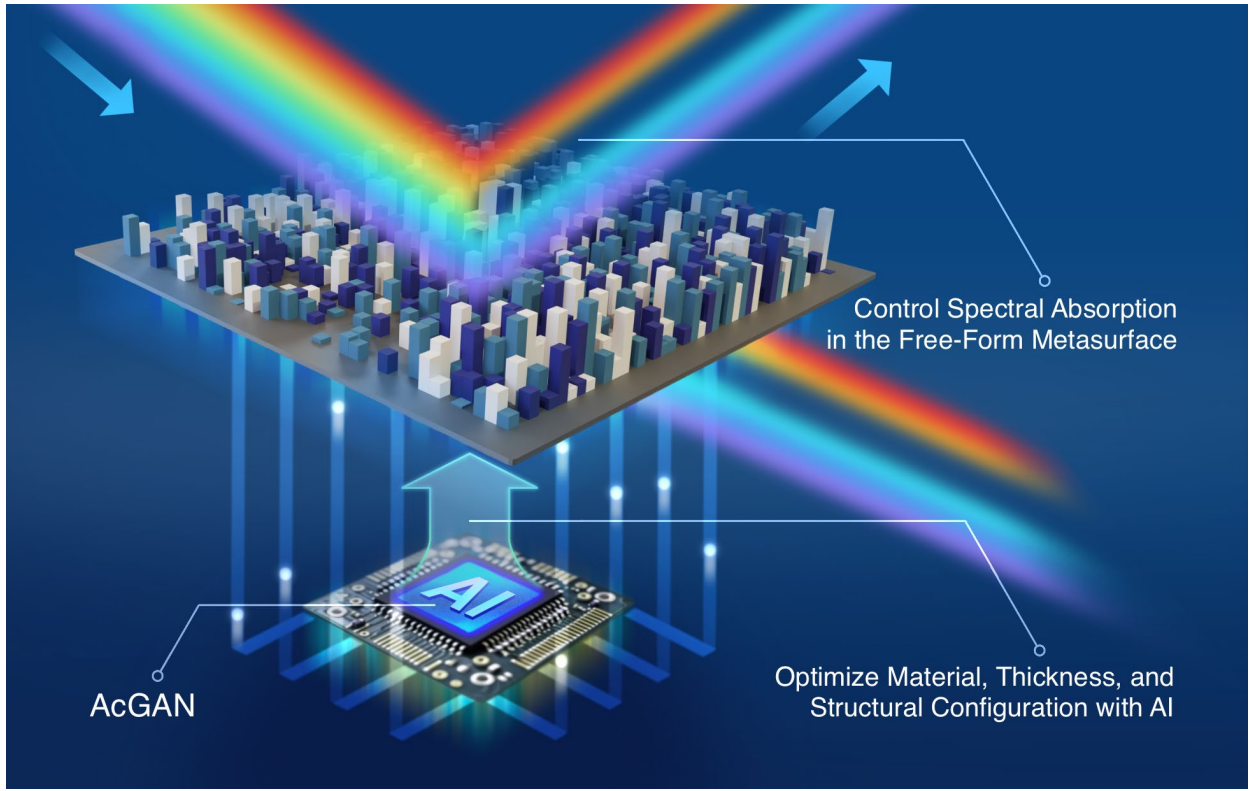

**Figure 1.** AI-enhanced free-form metasurface inverse design schematic, illustrating the application of AI to optimize material properties (indicated by color variations), meta-atom thickness (represented by height differences), and structural configuration (depicted through column arrangement) to precisely control and enhance spectral absorption.

**METHODS**

In this study, we address the complex task of designing free-form metasurface by proposing an advanced AI-driven framework named AcGAN, aiming to enhance both EM fidelity and structural diversity in metasurface designs. **Figure 2** outlines the AcGAN architecture, which includes four essential components: controller, generator, discriminator, and AnchorNet, each uniquely contributing to enhance EM fidelity and structural diversity of the generated metasurfaces. The process starts with the pre-trained AnchorNet predicting spectral properties, laying the groundwork for adversarial training. Initially, the discriminator is calibrated using precomputed control vectors that replace raw spectral data with structured inputs, streamlining the evaluation process. These inputs allow the discriminator to accurately assess the authenticity and spectral fidelity of the designs, ensuring they align with predefined criteria. Training then shifts focus to the generator, which is optimized through a specialized adversarial loss function to enhance its ability to produce structurally diverse and realistic metasurface designs. The training is iterative, with the generator and discriminator being refined alternately to ensure the designs meet the targeted spectral characteristics effectively. Detailed pseudocode is provided in the Section S1 (Supporting Information).

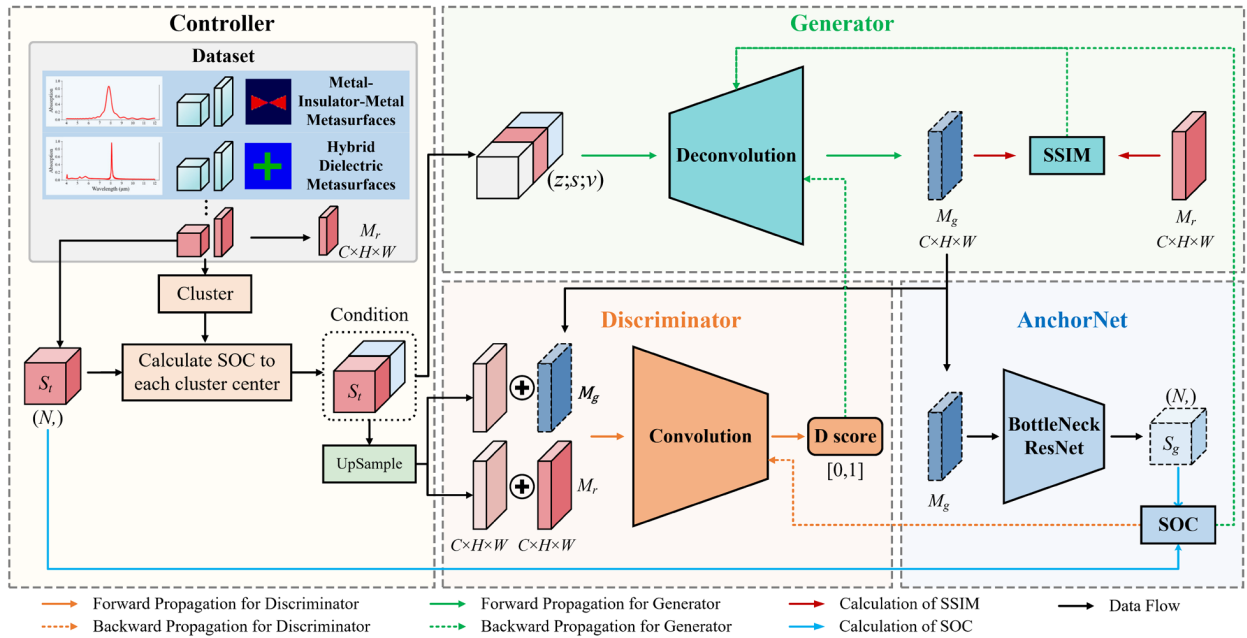

**Figure 2.** The architecture of AcGAN for metasurface design. The controller manages data clustering and processing, enhancing the structural diversity of designs by enabling the generator to explore various configurations. The generator creates metasurface designs based on these organized data inputs. The discriminator assesses the designs for authenticity and spectral fidelity, ensuring they meet predefined performance standards. AnchorNet guides both the generator and discriminator, providing real-time feedback to improve electromagnetic fidelity of the designs.

Building on the AcGAN framework, our approach further explores the engineering of metasurface parameters within a defined design space $M \in \mathbb{R}^d$ to align the spectral response $S \in \mathbb{R}^m$ of the metasurface with specified target spectra. Employing EM simulation tools such as Lumerical FDTD for forward mapping $F: M \to S$, we obtain reliable predictions of spectral responses for given metasurface configurations. The crux of the inverse design problem is to determine an optimal set of design parameters $m^*$ that minimizes the discrepancy between the spectrum $s_g$ of generated metasurface and the desired target spectrum $s_t$, represented by the optimization problem:

$$m^* = \underset{m \in M}{\mathrm{argmin}}\, \mathrm{Loss}(s_t, s_g), \tag{1}$$

where $s = F(m)$, and Loss quantifies the distance between $s_t$ and $s_g$. Additionally, the framework seeks to maximize the differences both between $M_t$ and $M_g$, as well as among multiple $M_g$ configurations, thereby promoting greater diversity in the metasurface designs. To ensure high electromagnetic fidelity in metasurface designs, we propose a novel spectral similarity metric named Spectral Overlap Coefficient (SOC), defined as follows:

$$SOC = 1 - \frac{\sum min(s_t, s_g)}{\sum max(s_t, s_g)}, \tag{2}$$

where $min(s_t, s_g)$ and $max(s_t, s_g)$ are computed element-wise across the spectral vectors $s_t$ and $s_g$. This formula quantitatively measures the extent of spectral overlap, providing a direct assessment of similarity that is especially useful for complex spectral features such as resonance peaks and specific absorption bands. A SOC value nearing zero signifies high similarity, which offers a direct and adaptable measure for spectral congruence, making it invaluable for evaluating and optimizing metasurface designs across diverse spectroscopic applications. SOC directly quantifies the extent of spectral overlap, offering a more granular and accurate measure of spectral congruence. This is particularly advantageous as it ensures a comprehensive alignment of all spectral features, critically assessing the match between peaks and troughs within the spectra. The adoption of SOC transforms our ability to design metasurfaces with high precision, aligning closely with specified EM requirements and surpassing the limitations of conventional design methodologies, as detailed in the Section S2 (Supporting Information).

In our inverse design framework, two types of metasurfaces are scrutinized, metal-insulator-metal (MIM) constructs characterized by broad Lorentzian absorption spectra arising from plasmonic resonances at the metal-dielectric interface[31]. These spectra are pivotal for applications demanding robust thermal emissivity and efficient photothermal energy conversion. Hybrid dielectric metasurfaces, wherein sub-wavelength cavity resonances yield Fano-resonant profiles, offering sharply defined spectral features optimal for discerning optical sensor technologies[32]. This bifurcation presents a complex challenge, necessitating a modeling approach capable of accommodating the significant spectral divergences inherent to each metasurface type. Following the approach outlined in Yeung's work[27], we encode two types of metasurfaces into $C \times H \times W$ pixel RGB images, where $C$ represents the number of color channels, $H$ denotes the image height, and $W$ indicates the image width. For MIM structures, the red channel encodes the plasma frequency $\omega_p$ of metal resonators, the blue channel denotes the dielectric layer's thickness ($d$) in nanometers, and the green channel is unused and set to zero. For hybrid dielectric structures, the green channel represents the real refractive index ($n$) of dielectric resonators, the blue channel continues to denote layer thickness ($d$) and the red channel is unused and set to zero. Additionally, the spectral $s$ range from 4-12 $\mu m$ is discretized into $N$ discrete points. This encoding strategy, while established, is crucial to AcGAN, enabling standardized representation of diverse metasurface types into a unified RGB format and facilitating the systematic discretization of spectral responses. The detailed encoding and decoding processes are shown in **Figure 3**.

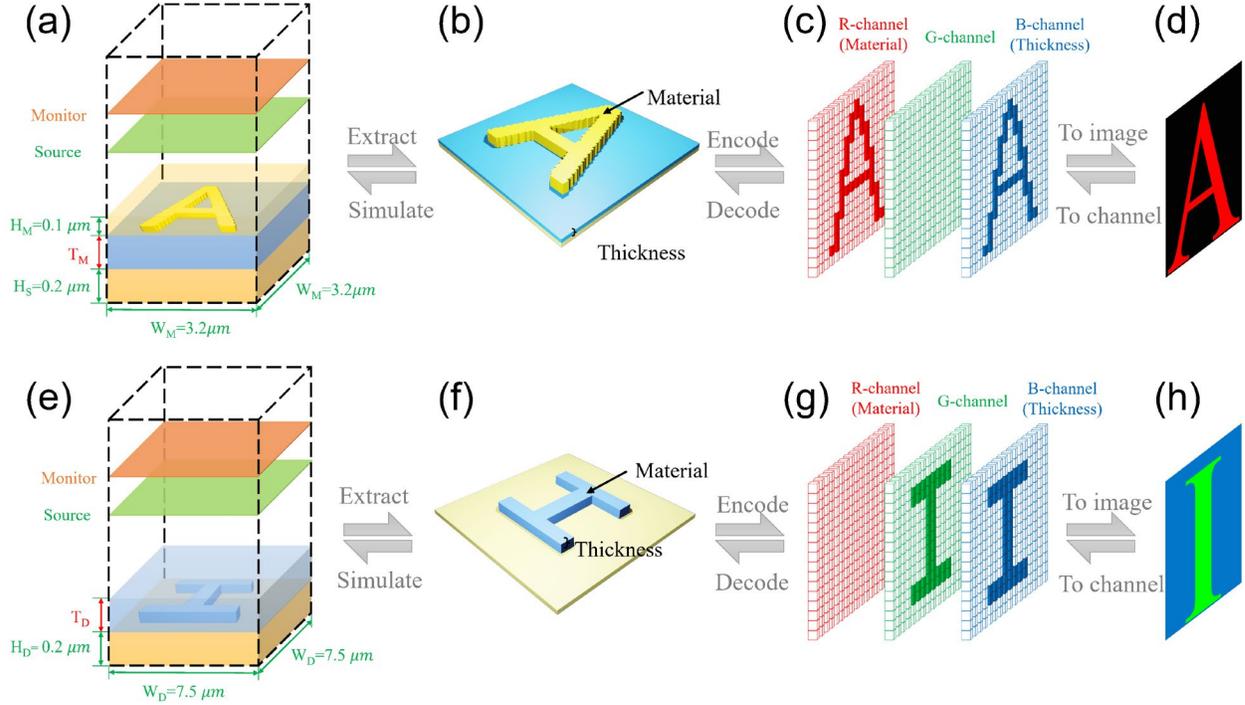

**Figure 3.** Schematic diagram of coding and decoding process of AcGAN. (a) and (e), the MIM structure (3.2 × 3.2 μm² unit cell) includes a 0.2 μm metal layer, a variable-height Al$_2$O$_3$ dielectric layer, and a freeform resonator of 0.1 μm height, while the hybrid structure (7.5 × 7.5 μm² unit cell) features a 0.2 μm metal layer with a dielectric freeform resonator of unspecified height. (b) and (f) show the 3D-rendered metasurfaces: the MIM resonators are composed of gold, silver, or aluminum with plasma frequencies of 1.91 PHz, 2.32 PHz, or 3.57 PHz, and dielectric layers of 100, 200, or 300 nm thickness; the hybrid resonators are made of ZnSe, Si, or Ge with refractive indices of 2.41, 3.42, or 4.01, and dielectric layers of 500, 750, or 950 nm thickness. In (c) and (g), the structures are encoded into 64×64×3 RGB images, where the R channel encodes material, the G channel encodes geometry, and the B channel encodes thickness. (d) and (h) present the final decoded images representing the letters "A" (MIM) and "I" (Hybrid).

Our framework employs an advanced controller mechanism to address the lack of structural diversity, which acts as an intelligent hub within our architecture, orchestrating the flow and preprocessing of spectral data through sophisticated clustering strategies. Specifically, we utilize the K-means clustering algorithm to segment the training spectral dataset $S = \{s_1, s_2, \ldots, s_n\}$ where each $s_i$ represents a unique spectral data point in $\mathbb{R}^m$. The algorithm partitions $S$ into $k$ distinct clusters, optimizing the following objective:

$$\min_{C} \sum_{i=1}^{k} \sum_{s \in C_i} \|s - c_i\|^2, \tag{3}$$

where $C = \{C_1, C_2, \ldots, C_k\}$ represents the set of clusters, and $c_i$ is the centroid or cluster center of $C_i$, embodying the average spectral profile of the cluster. These centroids are then used as reference points to compute the SOC for given spectral input, resulting in a $k$-dimensional vector $v = [\text{SOC}(s, c_1), \text{SOC}(s, c_2), \ldots, \text{SOC}(s, c_k)]$. This vector quantitatively describes the input's alignment with pre-identified spectral categories, enriching the input representation with both detailed and contextual spectral information. The resultant vector $v$ is then concatenated with the original spectral data $s$ to form a comprehensive control vector $u = [s; v]$, which is then input into the generator and discriminator. This enriched input empowers the generator to explore a wider design space, promoting the creation of diverse and functionally tailored metasurfaces. By integrating both detailed and aggregated spectral data, the controller ensures designs not only vary more broadly but also align closely with desired spectral characteristics, addressing the challenge of structural diversity problems.

Armed with the comprehensive control vector $u$—rich in both granularity and contextual insight—the generator is poised to harness this data for metasurface design. As depicted in **Figure** 4(a), the generator, equipped with a control vector $u$ and a latent vector $z$, leverages deconvolutional layers to transform enriched spectral data into precise spatial patterns. These layers, coupled with a noise vector for randomness, facilitate a broad exploration of design spaces—crucial for achieving one-to-many mappings in metasurface designs. The process is refined through up-sampling, which ensures the preservation of essential spectral features, allowing the generator to produce diverse and functionally effective metasurfaces. The generator's effectiveness is quantified by a loss function $L_G$, comprising three pivotal components:

$$L_G = \gamma L_{adv}^G + \alpha L_{\text{spectral}} + \beta L_{\text{structural}} \tag{4}$$

The adversarial loss component is defined as: $L_{adv}^G = -\mathbb{E}_{z \sim \mathbf{p}_Z(z), u \sim p_U(u)}[\log D(G(z,u), u)]$ where $\mathbb{E}_{z \sim \mathbf{p}_Z(z), u \sim p_U(u)}$ denotes the expectation over the distributions of latent vectors $z$ and control vectors $u$. Here, $G(z,u)$ is the generator's output for a given latent vector $z$ and control vectors $u$, where $D(G(z,u)$ represents the discriminator's assessment of how real or fake the generated metasurface. $L_{\text{spectral}} = \text{SOC}(s_g, s_t)$ measures the spectral similarity, ensuring the generated metasurface aligns with the target spectrum. The structural loss $L_{\text{structural}}$ is calculated using the Structural Similarity Index (SSIM)[33] between the generated and referenced samples: $L_{\text{structural}} = \text{SSIM}(M_g, M_r)$. This metric is critical for promoting structural diversity by encouraging the exploration of unique metasurface configurations. Parameters $\alpha$ and $\beta$ strategically balance the spectral and structural loss components within the loss function, tailoring the generator's output to meet specific operational demands while promoting structural diversity. This strategy guarantees that the generated designs not only effectively deceive the discriminator, demonstrating their realistic characteristics, but also accurately meet the targeted EM specifications and display significant structural diversity, thereby increasing their practical utility across various applications.

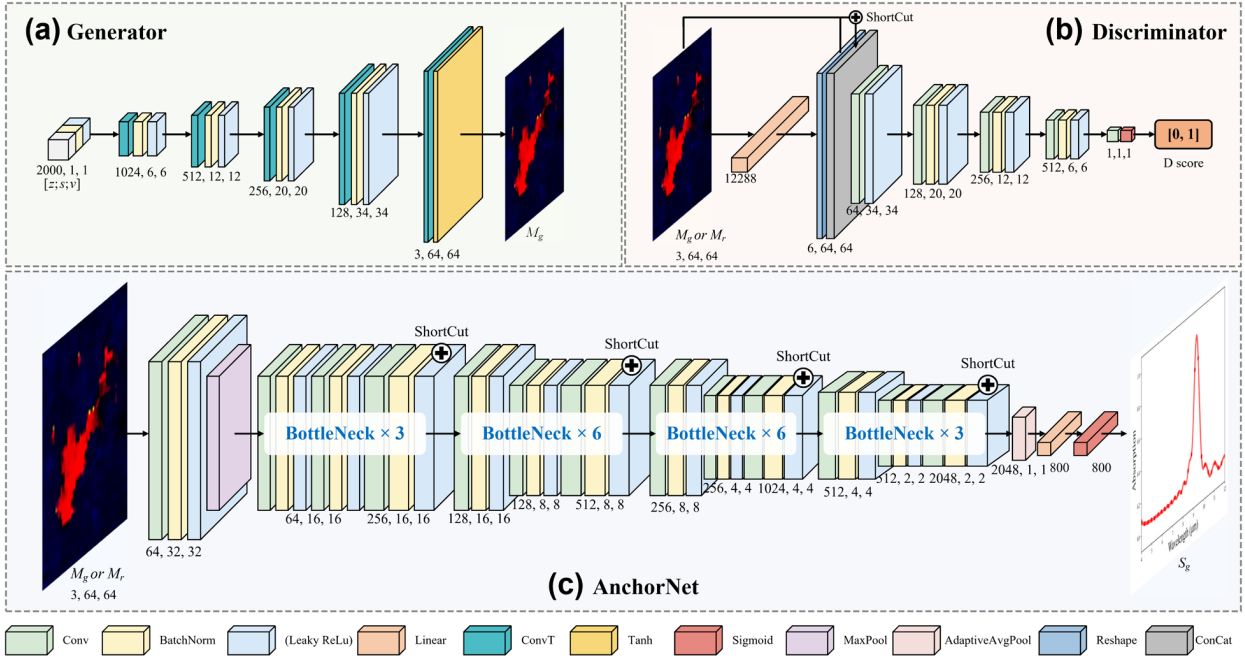

**Figure 4.** Detailed architectural overview of AcGAN components. (a) Generator: transforms control vectors $u$ into metasurface structures $M_g$ using deconvolutions; (b) Discriminator: assesses metasurface designs by applying convolutions to predict authenticity scores ranging from 0 to 1; (c) AnchorNet: predicts spectral responses $s_g$ from metasurface by leveraging bottleneck layers.

The discriminator, as depicted in **Figure 4**(b) is integral to the AcGAN framework, tasked primarily with validating the authenticity of the metasurface designs generated by the generator. It employs a sophisticated convolutional network to critically assess if the generated designs accurately reflect real metasurface EM properties. The discriminator assesses the quality of the synthesized designs using both learned features and heuristics from the training phase. It acts as the critical feedback component, thus guiding the generative process toward the production of metasurfaces with enhanced practical applicability. The discriminator's functionality is meticulously evaluated through a composite loss function $L_D$:

$$L_D = \gamma(L_{adv}^D + L_{\text{mismatch}}) + \alpha L_{\text{spectral}}, \qquad (5)$$

where the adversarial loss is expressed as: $L^D_{adv} = \mathbb{E}_{(M_r,s_t)\sim p_{\text{data}}(M_r,s_t)}[\log D(M_r, s_t)] - \mathbb{E}_{z\sim p_Z(z), u\sim p_U(u)}[\log(1 - D(G(z,u)))]$. This component compares the discriminator's predictions for referenced metasurface data $(M_r, s_t)$, sampled from the dataset, and generated data $G(z, u)$ from the generator. The first term, $E_{(M_r,s_t)\sim p_{\text{data}}(M_r,s_t)}[\log D(M_r, s_t)]$, encourages the discriminator to correctly identify real metasurfaces, maximizing the log probability of recognizing real designs. The second term, $E_{(z\sim p_Z(z), u\sim p_U(u))}[\log(1 - D(G(z,u), u))]$, penalizes the discriminator for falsely classifying generated metasurfaces as real, pushing it to distinguish between authentic and generated samples effectively. The mismatch loss, $L_{\text{mismatch}}$, enhances the discriminator's ability to detect inconsistencies between the referenced and generated metasurface, which is computed as:

$$L_{\text{mismatch}} = -\mathbb{E}_{z\sim p_Z(z), u'\sim p_U(u')}[\log(1 - D(G(z, u'), u))], \qquad (6)$$

where $u'$ represents a control vector that mismatches the intended input conditions for $z$, compelling the discriminator to reject generated metasurfaces that do not align with the required spectrum. We also use $L_{\text{spectral}}$ to further enhance the model's high EM fidelity capability.

Building upon the intricate interplay between the generator and discriminator within our AcGAN framework, AnchorNet emerges as a pivotal advancement, depicted in **Figure 4**(c). It incorporates a tailored BottleNeckResNet[34] architecture designed to predict the spectral responses of metasurface designs precisely. This module is finely tuned to minimize $SOC(\hat{s}, s)$, where $\hat{s}$ is the spectrum predicted by the AnchorNet, and $s$ is the ground truth spectrum obtained from EM simulations. An early stopping mechanism is integrated into the training protocol to halt the learning process after a predetermined number of epochs without improvement in training loss, ensuring computational efficiency and preventing overfitting. Integral to the AcGAN architecture,

AnchorNet revolutionizes metasurface design evaluation by providing rapid spectral response prediction capabilities. Specifically tailored to assess the EM performance of structures generated by AcGAN, it transcends the conventional visual assessment criteria used in CGANs. AnchorNet focuses is dedicated to aligning EM responses, enabling the design of metasurfaces optimized for high EM fidelity, independent of visual similarity to referenced designs.

The AcGAN framework integrates a carefully designed loss function that is crucial for balancing spectral precision and structural diversity in metasurface design. This loss function governs the interaction between the generator and discriminator to ensure that each design meets stringent spectral standards while exhibiting significant structural variation:

$$L = \min_G \max_D L(D, G) = \gamma(L_{adv} + L_{\text{mismatch}}) + \alpha L_{\text{spectral}} + \beta L_{\text{structural}}$$

$$= \gamma \begin{pmatrix} \mathbb{E}_{(M_r, s_t) \sim p_{\text{data}}(M_r, s_t), u \sim p_U(u)}[\log D(M_r, s_t, u)] + \\ \mathbb{E}_{z \sim p_Z(z), u \sim p_U(u)}[\log(1 - D(G(z, u), u))] - \\ \mathbb{E}_{z \sim p_Z(z), u' \sim p_U(u')}[\log(1 - D(G(z, u'), u))] \end{pmatrix} \quad (7)$$

$$+ \alpha \, \text{SOC}(s_g, s_t) + \beta \, \text{SSIM}(M_g, M_r)$$

Where adversarial loss $L_{adv}$ enables the generator to improve in deceiving the discriminator by making generated metasurfaces more realistic, the mismatch loss $L_{\text{mismatch}}$ ensures that the generated metasurface aligns with the control vectors, further enhancing design accuracy, the spectral loss ensures that the generated metasurfaces' spectral responses match the target spectra, and the structural loss encourages structural diversity by comparing the generated metasurfaces to referenced designs. This formulation ensures that each generated design meets stringent spectral criteria while achieving structural diversity, validated through comprehensive numerical simulations and material property analyses. AcGAN generates key parameters such as $\omega_p$, $d$ and $n$, facilitating the exploration of new materials and structures beyond traditional limitations, ultimately enhancing the efficiency and customization of metasurface designs.

**RESULTS**

To validate the innovative contributions of our AcGAN framework, we conducted a series of experiments focusing on key performance metrics such as spectral fidelity, structural diversity, and computational efficiency. Specifically, we tested AcGAN's ability to generate metasurface designs that meet precise spectral response criteria while overcoming the limitations of existing methods in structural diversity. By employing the novel SOC alongside traditional MSE, we quantitatively assessed the accuracy of the generated designs. Additionally, we explored AcGAN's capacity to generate diverse metasurface configurations for the same target spectrum, leveraging its one-time training advantage for rapid and efficient design iterations. All the experiments were conducted on a computational setup of Intel Xeon E5-2680 CPU (2.50GHz) and an NVIDIA GeForce RTX 3090 GPU with 24GB of VRAM, operating Python 3.9.12 on the Ubuntu Linux platform.

We firstly evaluated AnchorNet's performance within the AcGAN framework using a dataset of 18,768 metasurface structures, with one-third categorized as hybrid and two-thirds as MIM. To ensure thorough testing, 90% of the dataset was allocated for training, and the remaining 10% served as the test set; the detailed hyperparameter setting of AnchorNet is shown in the Section S3 (Supporting Information). As illustrated in **Figure 5**(a), SOC losses decreased progressively over 342 epochs, with training ceasing upon reaching an early stopping threshold of 30 consecutive epochs without validation loss improvement. The training consumed 136 minutes, with final SOC losses for training and testing converging to 0.0405 and 0.0807, respectively. After training, AnchorNet could predict the spectrum of metasurface in an average time of $3.6 \times 10^{-4}$ seconds, reducing computation time to approximately 1/1,600,000th of the 560.48 seconds required by FDTD simulations. This dramatic reduction in computational demand significantly accelerates the evaluation process for metasurface inverse design, enabling faster iterations and enhancements.

The assessment of AnchorNet across the dataset revealed differences between the hybrid and MIM categories, as detailed in **Figure 5**(b) to (i). The SOC distribution in **Figure 5**(b) and (c) indicates that hybrid structures predominantly appear in lower SOC intervals, suggesting AnchorNet more accurately predicts their complex spectral features. **Figure 5**(d) to (i) explicitly demonstrate AnchorNet's spectral prediction capabilities by presenting instances with the minimum, mean (close to the average SOC), and maximum SOC metrics. Specifically, **Figure 5**(d) to (f) for hybrid structures and **Figure 5**(g) to (i) for MIM structures illustrate the accuracy of spectral predictions through comparative plots that juxtapose predicted spectra with ground truth spectra. The improved accuracy for MIM structures is due to their simpler Lorentzian profiles, which are smoother and more predictable than the complex, asymmetric Fano resonances of hybrid structures. These Fano resonances, with sharp variations from quantum interference, present significant predictive challenges. Additionally, the spectral comparison in **Figure 5**(e) and 4(h) highlights a key discrepancy: despite a higher MSE in **Figure 5**(h), the spectral alignment closely matches the ground truth, particularly around peak regions. In contrast, **Figure 5**(e) shows significant deviations at peak intensities and across broader spectral regions, highlighting MSE's limitations as a reliable metric. SOC, by providing a more accurate measure of spectral similarity, proves superior to MSE in evaluating spectral fidelity. Inspird by Metric Learning[35], we evaluated MSE and SOC for spectral data dimensionality reduction. As detailed in Section S4 (Supporting Information), our analysis using Principal Component Analysis (PCA)[36], Locally Linear Embedding (LLE)[37], t-Distributed Stochastic Neighbor Embedding (t-SNE)[38], and Autoencoders (AE)[39] revealed MSE's limitations in distinguishing classes. In contrast, SOC significantly improved class separability, as shown in Figure S2, demonstrating its effectiveness in preserving intrinsic spectral properties crucial for advanced optical applications.

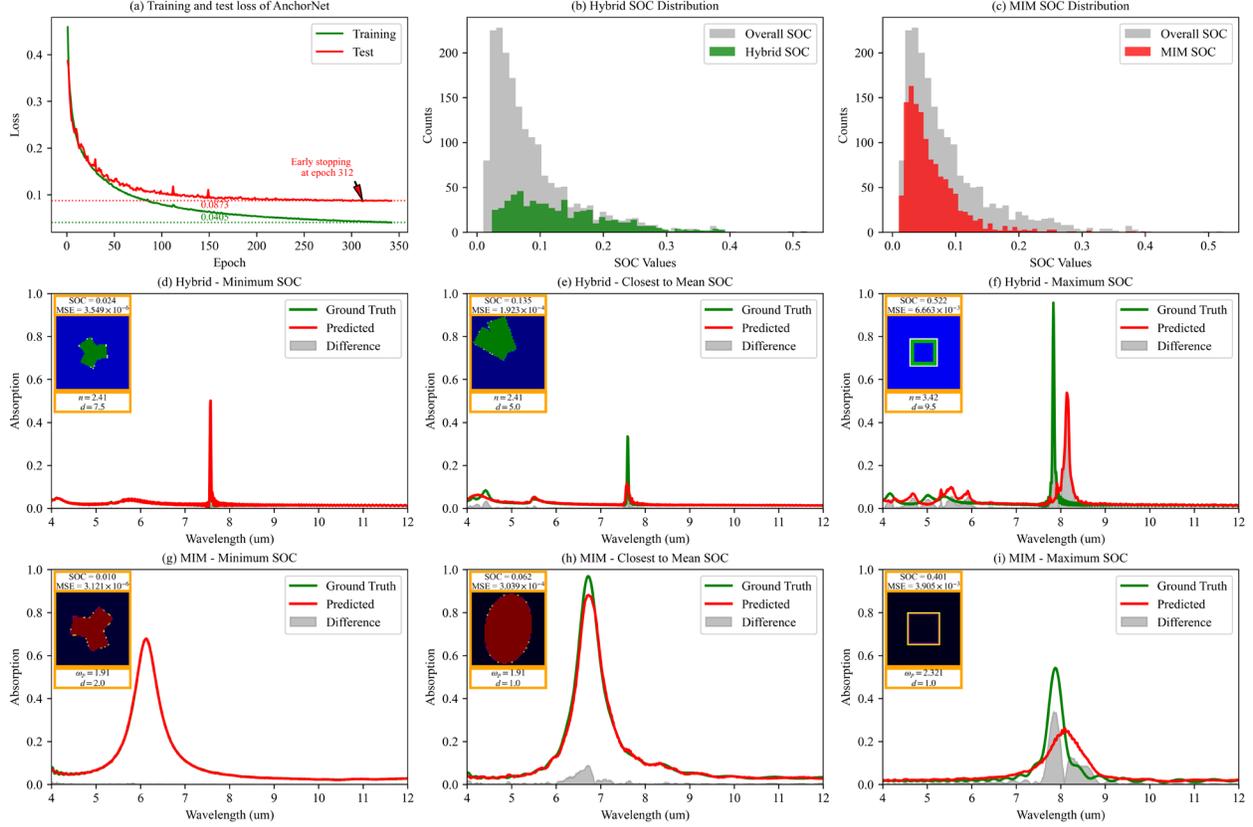

**Figure 5**. Performance evaluation and spectral predictions of AnchorNet. (a) Training and validation SOC loss curves with early stopping implementation. (b)-(c) SOC distributions: SOC distributions for hybrid (b) and MIM (c) metasurfaces indicating predictive accuracy. (d)-(i) Spectral comparisons for hybrid and MIM Structures: Displays minimum, mean, and maximum SOC scenarios for hybrid (d)-(f) and MIM (g)-(i) metasurfaces, with insets showing respective metasurface. Red and green lines represent actual and predicted spectra of AnchorNet, respectively.

To evaluate the performance of our AcGAN method against existing metasurface inverse design techniques, **Table 1** provides a comparison across key metrics including training and generation efficiency, MSE, and SOC. Detailed hyperparameter settings and analysis for AcGAN are provided in the Section S5 (Supporting Information). The hyperparameter settings for comparative methods were adopted from their original papers, and all methods were tested on the same dataset. The results are based on a random selection of 100 spectral data points from the test set. For each data point, corresponding metasurface structures were generated and then simulated using Lumerical FDTD to obtain what can be considered the ground truth spectrum of the designed structures. To address the challenge of limited structural diversity and assess the robustness of designs generated by AcGAN, 256 distinct latent vectors were employed for each target spectrum to explore variations in design accuracy and consistency. Thanks to CUDA's parallel computing capabilities, generating 256 metasurfaces takes nearly the same time as generating a single one. The design with the lowest SOC relative to the target spectrum is selected as the final design. Traditional physics-inspired methods, in contrast, demonstrate significant inefficiencies as they often require days to months to design a single metasurface and do not support one-time training. Heuristic algorithms like GA[40], PSO[41], and DE[42] are faster but achieve only moderate SOC and MSE, indicating suboptimal spectral fidelity. In contrast, AI-based techniques such as DNN[43], VAE[44], and CGAN[27] significantly reduce prediction latency to milliseconds while improving SOC and MSE. Among these, our AcGAN method excels by recording the fastest generation time of only $4.2 \times 10^{-4}$ seconds, and achieving the lowest MSE ($1.120 \times 10^{-3}$) and SOC (0.139). Additionally, the one-time training feature of machine learning methods presents considerable advantages over the iterative, resource-intensive nature of traditional and heuristic approaches.

**Table 1.** Comparison of metasurface design methods based on training and design efficiency, MSE, and SOC: "Training time" indicates the time required for model training, while "Generation time" measures the time needed to design a metasurface that matches the target spectrum. MSE and SOC assess the spectral fidelity of the designed metasurface relative to the desired spectrum. "One-time training" indicates whether the model requires retraining for new spectral targets.

| Method | Training Time | Generation Time | MSE | SOC | One-Time Training |
|---|---|---|---|---|---|
| Physics-inspired | ------ | Days or Months | ------ | ------ | No |
| GA[40] | ------ | 3.89 hours | $2.120 \times 10^{-2}$ | 0.564 | No |
| PSO[41] | ------ | 2.23 hours | $2.010 \times 10^{-2}$ | 0.557 | No |
| DE[42] | ------ | 4.45 hours | $2.070 \times 10^{-2}$ | 0.572 | No |
| DNN[43] | 13.4 hours | $6.1 \times 10^{-4}$ s | $1.500 \times 10^{-2}$ | 0.471 | Yes |
| VAE[44] | 16.8 hours | $6.4 \times 10^{-4}$ s | $1.450 \times 10^{-2}$ | 0.454 | Yes |
| CGAN[27] | 9.62 hours | $4.2 \times 10^{-4}$ s | $4.151 \times 10^{-3}$ | 0.274 | Yes |
| **AcGAN** | **4.16 hours** | $\mathbf{4.2 \times 10^{-4}}$ **s** | $\mathbf{1.120 \times 10^{-3}}$ | **0.139** | **Yes** |

**Figure 6** presents 9 representative cases from the test dataset, each showcasing the design generated by AcGAN with the lowest SOC relative to the target spectrum, highlighting the model's ability to achieve high spectral fidelity. The spectra from FDTD simulations of metasurfaces designed by AcGAN closely align with the target spectra, underscoring AcGAN's unprecedented efficiency and accuracy in generating metasurfaces precisely tailored to specific spectral requirements, a significant advancement over traditional methods. Moreover, the close match between the spectra from referenced metasurface structures and the target spectra predicted by AnchorNet emphasizes AnchorNet's exceptional predictive capability, which is critical for ensuring high electromagnetic fidelity in the design process. Notably, AcGAN accurately designed MIM structures for Lorentzian spectra and hybrid structures for Fano spectra, successfully distinguishing between different physical mechanisms during training without confusion. This demonstrates AcGAN's innovative capability to effectively differentiate between distinct metasurface types and their corresponding absorption spectra, showcasing its proficiency not only in achieving high design accuracy but also in understanding and applying different physical mechanisms—a significant advancement in metasurface design methodologies.

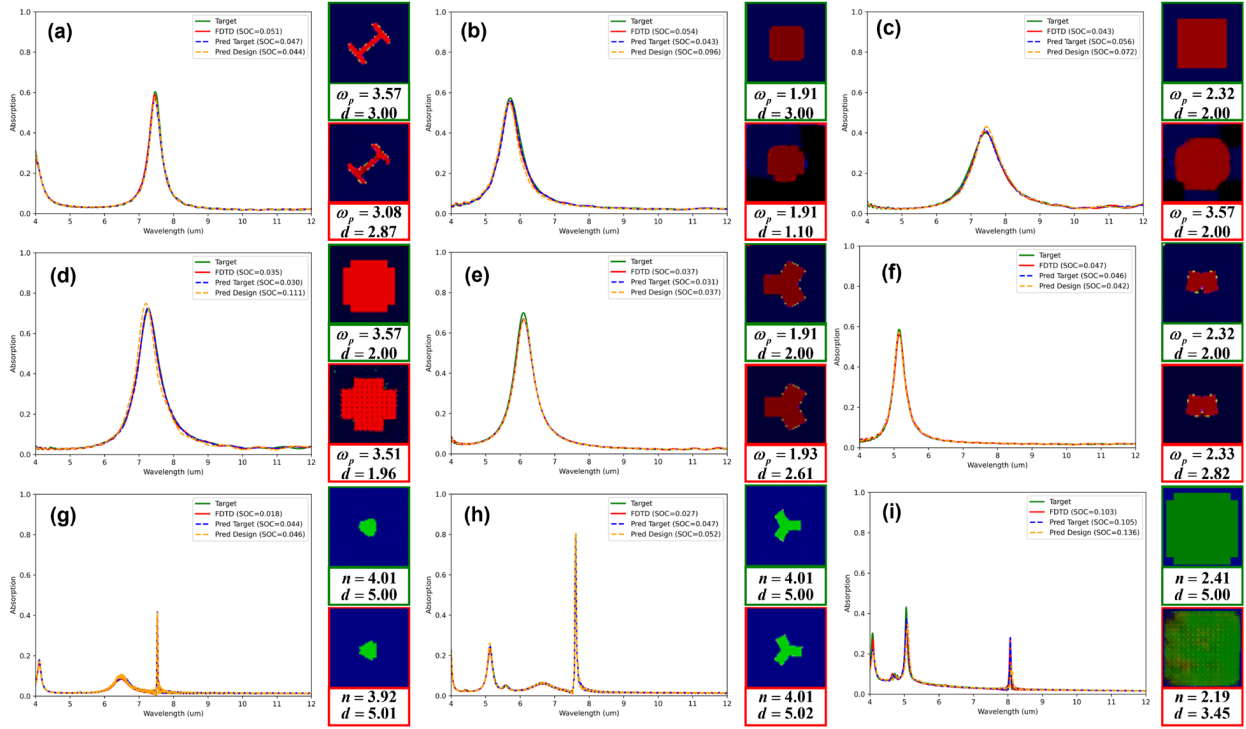

**Figure 6.** Randomly selected spectra from the test dataset served as input targets for AcGAN. This figure presents a comparative analysis between the target spectra (solid green lines) and the corresponding spectra of metasurfaces designed by AcGAN after FDTD simulation (solid red lines). Dashed blue and yellow lines represent the spectra predicted by AnchorNet for referenced metasurface and AcGAN-designed metasurfaces, respectively. To the right of each plot, green-framed images depict reference metasurface structures with material and thickness parameters, while red-framed images show AcGAN-designed metasurfaces, annotated with material types and thicknesses ($t$ in nanometers), and plasma frequency ($\omega_P$) values are given in PHz.

AcGAN demonstrates a notable capacity to design metasurfaces that closely match target spectra while significantly diverging in structural dimensions, material properties, and dielectric thickness, as shown in **Figure 6** and **Figure 7**. Analysis of the three metasurfaces with the lowest SOC reveals that, although their absorption spectra closely match the target, their physical configurations vary significantly. For instance, material properties deviate by an average of 12.9%, and dielectric thicknesses vary by 20.0%, highlighting the model's ability to achieve diverse designs beyond traditional visual constraints. In particular, **Figure 7** (b) highlights that design1's thickness is reduced by 26.7% compared to the referenced metasurface, simplifying the manufacturing process. Despite these variations, the average SSIM between the generated and referenced metasurfaces is 0.727, indicating substantial structural differences while maintaining functional integrity, as shown in **Figure 7** (b) and (c).

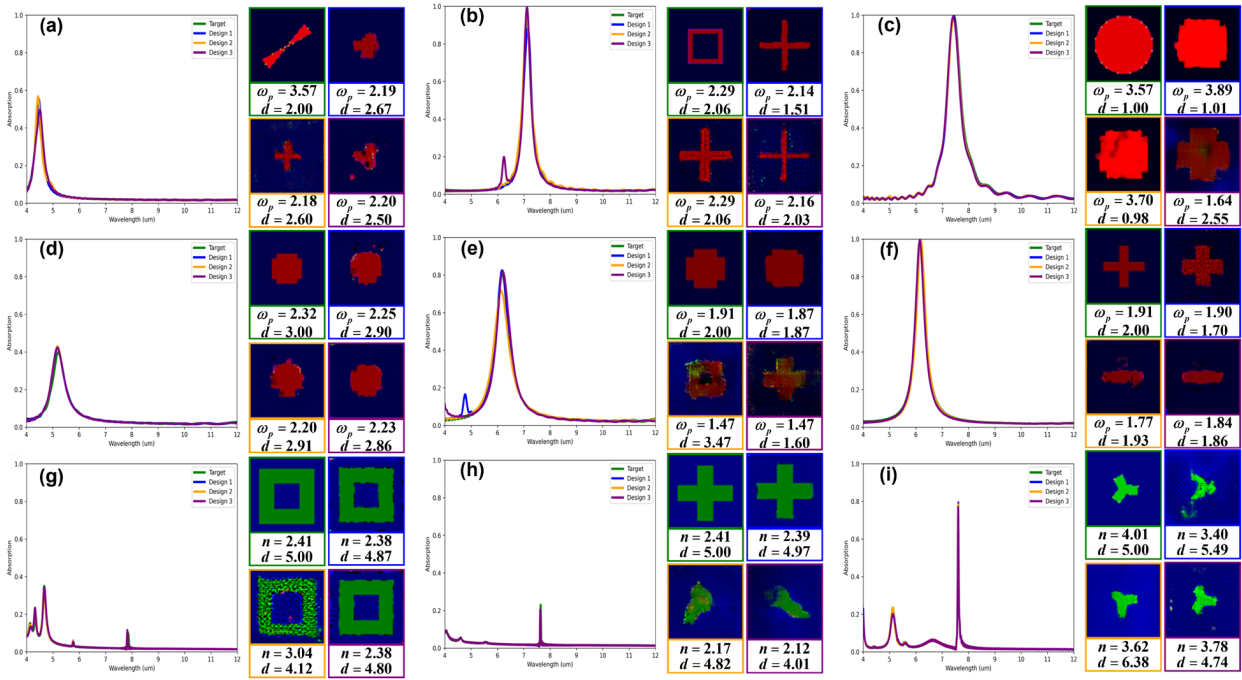

**Figure 7.** Demonstrating AcGAN's capability for design diversity: This figure compares target spectra (solid green lines) with three distinct spectra generated by AcGAN(solid blue, yellow, and purple lines), highlighting AcGAN's ability to create multiple diverse metasurface designs from a single target specification. Adjacent to each spectral plot, images within green frames depict referenced metasurface structures with detailed material and thickness parameters. Corresponding images in blue, yellow, and purple frames showcase the various designs created by AcGAN, each annotated with specific material types, thicknesses and plasma frequency values.

To further illustrate AcGAN's ability to enhance design diversity, **Figure 8** presents the near-field electric distributions in the XY plane for both MIM and hybrid metasurface structures. This visualization demonstrates AcGAN's ability to not only match the target spectral responses but also innovate in the spatial arrangement of meta-atoms across various planes. The corresponding electric field variations in the XZ and YZ planes, which exhibit similarly diverse distributions, are discussed in Section S6 (Supporting Information). The design versatility enabled by AcGAN allows a single imaging system to perform multiple functions, such as standard and polarimetric imaging, without requiring changes to the optical components. This adaptability significantly enhances the visualization of cellular or tissue structures across different depths and orientations. By generating metasurfaces with tailored EM functionalities, AcGAN expands the operational flexibility and efficiency of imaging systems, paving the way for broader applications.

To assess AcGAN's ability to handle arbitrarily-defined spectral challenges, we explored four spectral types: Fano, Lorentzian, Gate, and Gaussian. These spectra were generated according to the details in the Section S7 (Supporting Information), ensuring no same data in the training dataset. For each type, we generated two spectra and simulated 256 metasurface structures per spectrum. The results are shown in Section S8 (Supporting Information), each panel in Figure S5 contrasts the simulated spectra with the target, highlighting discrepancies in shaded areas and quantifying them with SOC and MSE values. The results show that while AcGAN closely approximates the true spectral characteristics for Fano and Lorentzian resonances (Figure S5 (a-d)), it exhibits deviations in Gate and Gaussian profiles, particularly at the spectral tails (Figure S5 (e-h)). This variance suggests AcGAN's robust performance on spectra present in the database but highlights the need for better model generalization to accommodate theoretically defined but unrepresented spectral types in the training set.

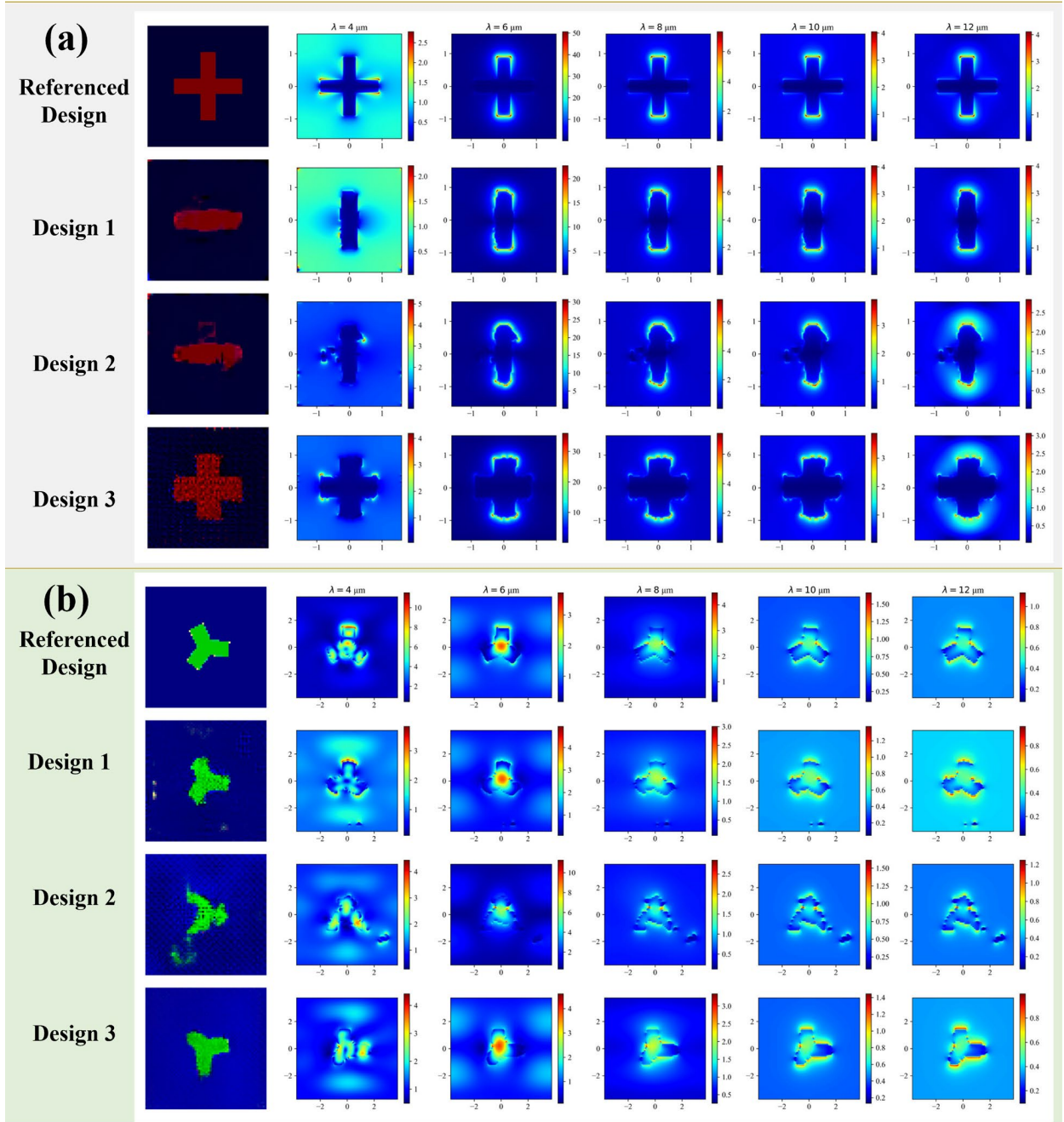

**Figure 8**. Near-field electric responses in the XY plane for MIM and hybrid metasurface: (a) MIM metasurface: Showcases the near-field electric responses at various wavelengths (4 *μm* to 12 *μm*) for MIM metasurface in **Figure 7** (f). The depth of the colors i reflects the magnitude of the near-field electric field strength. (b) Hybrid metasurface: Presents the near-field electric responses at wavelengths from 4 *μm* to 12 *μm* for hybrid metasurface in **Figure 7** (i).

**DISCUSSION**

To further analyze the performance of AcGAN for a variety of spectral designs and their relation to the training dataset, we propose the "weighted distance" metric which is defined as $D_w(s) = \sum_{i=1}^{k} \left( \frac{n_i}{N} \cdot \text{SOC}(s, c_i) \right)$, this metric quantifies the deviation of a spectrum $s$ from the cluster centroids $\{c_1, c_2, \ldots, c_k\}$ based on the training data. Here, $n_i$ is the number of spectra in the $i$-th cluster, and $N$ is the total number of spectra across all clusters. For empirical analysis, we generated 1,000 spectra for each of the four spectral types and calculated their respective $D_w$. Each spectrum was categorized into intervals of 0.05 in weighted distance, with ten spectra sampled per interval to ensure uniform coverage. The results are shown in Section S9 (Supporting Information), Figure S6 illustrates the correlation between SOC and weighted distance, with ellipses highlighting the general distribution of SOC against weighted distance for each type. Notably, spectra with lower weighted distances typically achieved lower SOC, indicating closer approximation to the target spectral characteristics. This trend was especially pronounced for Lorentzian and Fano spectra. Conversely, the Gate and Gaussian spectra demonstrated lower SOC, underscoring potential challenges in generating these spectra types due to their underrepresentation in the training data. The observed variance in AcGAN's performance across spectral types highlights the critical importance of training dataset diversity for model generalization. Expanding the dataset to include a broader spectrum of metasurface configurations could enhance the model's capability to accurately generate designs for a wider array of spectral responses. Future research will focus on augmenting the dataset and refining model algorithms to improve performance across less represented spectra, thereby broadening the practical applications of the AcGAN framework in metasurface design.

To thoroughly evaluate the AcGAN's robustness and adaptability to varying configurations, we executed a series of ablation studies, the detailed results of which are presented in the Section S10 (Supporting Information). Our ablation analyses underscored the pivotal roles of the cluster controller and AnchorNet's integration within both the generator and discriminator. The presence of the cluster controller led to a notable decrease in MSE and SOC, indicating higher fidelity EM design. Similarly, enabling AnchorNet in both model components significantly enhanced the fidelity of generated metasurface designs. Adjustments in the adversarial training dynamics, particularly variations in the k-value, defined as the number of generator updates per discriminator update, revealed that a balanced approach is crucial for stable training and optimal model performance. The studies highlighted that a k-value of 2 was optimal, significantly reducing MSE and increasing SOC compared to other configurations. Moreover, the impact of initial spectral loss weighting, represented by the parameter $\alpha$, was profound. Surprisingly, a lower weight ($\alpha = 0.1$) unexpectedly yielded better performance, suggesting that an excessive initial emphasis on spectral fidelity might hinder the model's ability to generalize across a broader design space. This finding points to the need for a balanced loss function that adequately emphasizes both spectral fidelity and adversarial robustness. Investigations into the effects of latent space dimensions and batch sizes further refined our understanding of model behavior. Optimal latent dimensions and smaller batch sizes tended to improve the model's precision and stability, indicating that finer granularity in the generation process aids in capturing the nuances of metasurface designs. Collectively, these findings underscore the intricate interdependencies within the AcGAN architecture and highlight the need for careful calibration to maximize performance. These insights are crucial for refining the AcGAN framework to enhance its practical applicability across diverse metasurface design scenarios, thereby extending the capabilities of current computational photonic design methods.

**CONCLUSION**

AcGAN marks a significant advancement in the inverse design of metasurfaces, showcasing an unprecedented combination of high electromagnetic fidelity and extensive structural diversity. This innovative framework, through rigorous empirical testing, has proven to effectively minimize the MSE by 73% compared to existing GAN approaches, demonstrating its robust capability to meet stringent spectral demands with enhanced precision. Crucially, AcGAN addresses the perennial challenges of electromagnetic fidelity and structural diversity that have impeded prior generative models. By integrating the SOC and AnchorNet, our framework not only assesses but significantly improves spectral fidelity, ensuring that each generated design adheres closely to the desired electromagnetic characteristics. This precision is vital for applications in complex optoelectronic systems, where exact spectral properties are critical for functionality. Furthermore, AcGAN innovatively incorporates a cluster-guided controller, which refines input processing and facilitates the exploration of diverse structural configurations. This feature is essential for overcoming the one-to-many mapping dilemma inherent in metasurface design, allowing for a broader range of functional possibilities within a single design process. The dynamic loss function, shifting focus from data-driven learning to optimizing spectral and structural outcomes, further underscores our method's adaptability and efficiency.

Looking ahead, future enhancements to AcGAN will focus on several critical areas: 1. **Optimizing AnchorNet**: Enhancing its predictive accuracy, particularly for hybrid structures with complex Fano resonance profiles, which currently present significant challenges. 2. **Enhancing encoding techniques**: Expanding the range of design variables to include more intricate and functional metasurface configurations, thereby broadening the scope of metasurface applications. 3. **Expanding dataset diversity**: Incorporating a broader array of metasurface structures, especially free-form designs, to improve the model's generalization capabilities and robustness. **4. Considering manufacturability and fabrication tolerances**: Addressing potential challenges in manufacturing complex metasurface designs by ensuring that the generated structures are not only diverse but also feasible to produce with existing fabrication technologies. This will improve the practical applicability of AcGAN-generated designs in real-world scenarios.

AcGAN represents not only a methodological breakthrough but also a scalable and robust framework that significantly accelerates the design process in nanophotonic applications. This makes AcGAN a pivotal tool for advancing operational metasurfaces, meeting the evolving demands of optoelectronics and related industries.

**ASSOCIATED CONTENT**

The Supporting Information is available in the Supporting Information file. Including the pseudocode of the training of AcGAN, literature review on spectral similarity evaluation metrics, detailed hyperparameter setting of AnchorNet, evaluating the impact of spectral similarity evaluation metrics on spectral data dimensionality reduction, detailed hyperparameter setting of AcGAN, diverse near-field electric responses of metasurfaces with similar absorption spectra designed by AcGAN, generation of four arbitrarily-defined spectra, performance of AcGAN of arbitrarily-defined spectrum, correlation between weighted distance and SOC for various spectral types, and ablation study.

**AUTHOR INFORMATION**

**Corresponding Author:** Xin Jin 一 Shenzhen International Graduate School, Tsinghua University, Shenzhen 518055, China; ORCID: 0000-0001-6655-3888; Email: jin.xin@sz.tsinghua.edu.cn;

Hongkun Cao一Peng Cheng Laboratory, Shenzhen 518055, ORCID: 0000-0002-3524-8570; China; Email: caohk@pcl.ac.cn

**Authors:** Yunhui Zeng一Shenzhen International Graduate School, Tsinghua University, Shenzhen 518055, China；Peng Cheng Laboratory, Shenzhen 518055, China

**Author Contributions:** Y.Z. conceived the idea and implemented the inverse design algorithm. X.J. and H.C. oversaw the research. The manuscript was written through the contributions of all authors. All authors have given approval to the final version of the manuscript.

**Notes:** The authors declare no competing financial interest.

## ACKNOWLEDGMENTS

This work is supported in part by Natural Science Foundation of China under Grant 62131011; in part by Shenzhen Project under Grant JSGG20220831095602005 and in part by the Major Key Project of PCL under Grant PCL2023A10-3.

## REFERENCES

(1) Chen, J.; Qian, C.; Zhang, J.; Jia, Y.; Chen, H. Correlating Metasurface Spectra with a Generation-Elimination Framework. *Nat Commun* **2023**, *14* (1), 4872. https://doi.org/10.1038/s41467-023-40619-w.
(2) Rana, A. S.; Zubair, M.; Danner, A.; Mehmood, M. Q. Revisiting Tantalum Based Nanostructures for Efficient Harvesting of Solar Radiation in STPV Systems. *Nano Energy* **2021**, *80*, 105520. https://doi.org/10.1016/j.nanoen.2020.105520.
(3) Dai, J. Y.; Zhao, J.; Cheng, Q.; Cui, T. J. Independent Control of Harmonic Amplitudes and Phases via a Time-Domain Digital Coding Metasurface. *Light Sci Appl* **2016**, *7* (1), 90. https://doi.org/10.1038/s41377-018-0092-z.
(4) Wang, L.; Kruk, S.; Tang, H.; Li, T.; Kravchenko, I.; Neshev, D. N.; Kivshar, Y. S. Grayscale Transparent Metasurface Holograms. *Optica* **2016**, *3* (12), 1504. https://doi.org/10.1364/OPTICA.3.001504.
(5) Chen, X.; Zhang, Y.; Huang, L.; Zhang, S. Ultrathin Metasurface Laser Beam Shaper. *Advanced Optical Materials* **2014**, *2* (10), 978–982. https://doi.org/10.1002/adom.201400186.

(6) Khorasaninejad, M.; Chen, W. T.; Devlin, R. C.; Oh, J.; Zhu, A. Y.; Capasso, F. Metalenses at Visible Wavelengths: Diffraction-Limited Focusing and Subwavelength Resolution Imaging. *Science* **2016**, *352* (6290), 1190–1194. https://doi.org/10.1126/science.aaf6644.

(7) Balthasar Mueller, J. P.; Rubin, N. A.; Devlin, R. C.; Groever, B.; Capasso, F. Metasurface Polarization Optics: Independent Phase Control of Arbitrary Orthogonal States of Polarization. *Phys. Rev. Lett.* **2017**, *118* (11), 113901. https://doi.org/10.1103/PhysRevLett.118.113901.

(8) Arbabi, A.; Horie, Y.; Bagheri, M.; Faraon, A. Dielectric Metasurfaces for Complete Control of Phase and Polarization with Subwavelength Spatial Resolution and High Transmission. *Nature Nanotech* **2015**, *10* (11), 937–943. https://doi.org/10.1038/nnano.2015.186.

(9) Wang, Z.; Li, T.; Soman, A.; Mao, D.; Kananen, T.; Gu, T. On-Chip Wavefront Shaping with Dielectric Metasurface. *Nat Commun* **2019**, *10* (1), 3547. https://doi.org/10.1038/s41467-019-11578-y.

(10) Jin, Z.; Janoschka, D.; Deng, J.; Ge, L.; Dreher, P.; Frank, B.; Hu, G.; Ni, J.; Yang, Y.; Li, J.; Yu, C.; Lei, D.; Li, G.; Xiao, S.; Mei, S.; Giessen, H.; zu Heringdorf, F. M.; Qiu, C.-W. Phyllotaxis-Inspired Nanosieves with Multiplexed Orbital Angular Momentum. *eLight* **2021**, *1* (1), 5. https://doi.org/10.1186/s43593-021-00005-9.

(11) Lin, R. J.; Su, V.-C.; Wang, S.; Chen, M. K.; Chung, T. L.; Chen, Y. H.; Kuo, H. Y.; Chen, J.-W.; Chen, J.; Huang, Y.-T.; Wang, J.-H.; Chu, C. H.; Wu, P. C.; Li, T.; Wang, Z.; Zhu, S.; Tsai, D. P. Achromatic Metalens Array for Full-Colour Light-Field Imaging. *Nat. Nanotechnol.* **2019**, *14* (3), 227–231. https://doi.org/10.1038/s41565-018-0347-0.

(12) Kim, G.; Kim, S.; Kim, H.; Lee, J.; Badloe, T.; Rho, J. Metasurface-Empowered Spectral and Spatial Light Modulation for Disruptive Holographic Displays. *Nanoscale* **2022**, *14* (12), 4380–4410. https://doi.org/10.1039/D1NR07909C.

(13) Liang, Y.; Lin, H.; Koshelev, K.; Zhang, F.; Yang, Y.; Wu, J.; Kivshar, Y.; Jia, B. Full-Stokes Polarization Perfect Absorption with Diatomic Metasurfaces. *Nano Lett.* **2021**, *21* (2), 1090–1095. https://doi.org/10.1021/acs.nanolett.0c04456.

(14) Chen, M.-H.; Chen, B.-W.; Xu, K.-L.; Su, V.-C. Wide-Angle Optical Metasurface for Vortex Beam Generation. *Nanomaterials* **2016**, *13* (19), 2680. https://doi.org/10.3390/nano13192680.

(15) Yu, Z.; Li, H.; Zhao, W.; Huang, P.-S.; Lin, Y.-T.; Yao, J.; Li, W.; Zhao, Q.; Wu, P. C.; Li, B.; Genevet, P.; Song, Q.; Lai, P. High-Security Learning-Based Optical Encryption Assisted by Disordered Metasurface. *Nat Commun* **2024**, *15* (1), 2607. https://doi.org/10.1038/s41467-024-46946-w.

(16) Wu, Y.; Chen, J.; Wang, Y.; Yuan, Z.; Huang, C.; Sun, J.; Feng, C.; Li, M.; Qiu, K.; Zhu, S.; Zhang, Z.; Li, T. Tbps Wide-Field Parallel Optical Wireless Communications Based on a Metasurface Beam Splitter. *Nat Commun* **2024**, *15* (1), 7744. https://doi.org/10.1038/s41467-024-52056-4.

(17) Li, Z.; Pestourie, R.; Lin, Z.; Johnson, S. G.; Capasso, F. Empowering Metasurfaces with Inverse Design: Principles and Applications. *ACS Photonics* **2022**, *9* (7), 2178–2192. https://doi.org/10.1021/acsphotonics.1c01850.

(18) Zhang, X.; Liu, Y.; Han, J.; Kivshar, Y.; Song, Q. Chiral Emission from Resonant Metasurfaces. *Science* **2022**, *377* (6611), 1215–1218. https://doi.org/10.1126/science.abq7870.

(19) Li, Z.; Kim, M.-H.; Wang, C.; Han, Z.; Shrestha, S.; Overvig, A. C.; Lu, M.; Stein, A.; Agarwal, A. M.; Lončar, M.; Yu, N. Controlling Propagation and Coupling of Waveguide

Modes Using Phase-Gradient Metasurfaces. *Nature Nanotech* **2017**, *12* (7), 675–683. https://doi.org/10.1038/nnano.2017.50.

(20) Maguid, E.; Yulevich, I.; Yannai, M.; Kleiner, V.; L Brongersma, M.; Hasman, E. Multifunctional Interleaved Geometric-Phase Dielectric Metasurfaces. *Light Sci Appl* **2017**, *6* (8), e17027–e17027. https://doi.org/10.1038/lsa.2017.27.

(21) Wang, S.; Wu, P. C.; Su, V.-C.; Lai, Y.-C.; Hung Chu, C.; Chen, J.-W.; Lu, S.-H.; Chen, J.; Xu, B.; Kuan, C.-H.; Li, T.; Zhu, S.; Tsai, D. P. Broadband Achromatic Optical Metasurface Devices. *Nat Commun* **2017**, *8* (1), 187. https://doi.org/10.1038/s41467-017-00166-7.

(22) Balthasar Mueller, J. P.; Rubin, N. A.; Devlin, R. C.; Groever, B.; Capasso, F. Metasurface Polarization Optics: Independent Phase Control of Arbitrary Orthogonal States of Polarization. *Phys. Rev. Lett.* **2017**, *118* (11), 113901. https://doi.org/10.1103/PhysRevLett.118.113901.

(23) Fu, Y.; Zhou, X.; Yu, Y.; Chen, J.; Wang, S.; Zhu, S.; Wang, Z. Unleashing the Potential: AI Empowered Advanced Metasurface Research. *Nanophotonics* **2024**, *13* (8), 1239–1278. https://doi.org/10.1515/nanoph-2023-0759.

(24) So, S.; Mun, J.; Park, J.; Rho, J. Revisiting the Design Strategies for Metasurfaces: Fundamental Physics, Optimization, and Beyond. *Advanced Materials* **2023**, *35* (43), 2206399. https://doi.org/10.1002/adma.202206399.

(25) Chen, W. T.; Zhu, A. Y.; Sanjeev, V.; Khorasaninejad, M.; Shi, Z.; Lee, E.; Capasso, F. A Broadband Achromatic Metalens for Focusing and Imaging in the Visible. *Nature Nanotech* **2018**, *13* (3), 220–226. https://doi.org/10.1038/s41565-017-0034-6.

(26) Chen, M. K.; Liu, X.; Sun, Y.; Tsai, D. P. Artificial Intelligence in Meta-Optics. *Chem. Rev.* **2022**, *122* (19), 15356–15413. https://doi.org/10.1021/acs.chemrev.2c00012.

(27) Yeung, C.; Tsai, R.; Pham, B.; King, B.; Kawagoe, Y.; Ho, D.; Liang, J.; Knight, M. W.; Raman, A. P. Global Inverse Design across Multiple Photonic Structure Classes Using Generative Deep Learning. *Advanced Optical Materials* **2021**, *9* (20), 2100548. https://doi.org/10.1002/adom.202100548.

(28) Yeung, C.; Pham, B.; Tsai, R.; Fountaine, K. T.; Raman, A. P. DeepAdjoint: An All-in-One Photonic Inverse Design Framework Integrating Data-Driven Machine Learning with Optimization Algorithms. *ACS Photonics* **2022**. https://doi.org/10.1021/acsphotonics.2c00968.

(29) Mirza, M.; Osindero, S. Conditional Generative Adversarial Nets. arXiv November 6, 2014. http://arxiv.org/abs/1411.1784 (accessed 2022-08-23).

(30) Baucour, A.; Kim, M.; Shin, J. Data-Driven Concurrent Nanostructure Optimization Based on Conditional Generative Adversarial Networks. *Nanophotonics* **2022**, *11* (12), 2865–2873. https://doi.org/10.1515/nanoph-2022-0005.

(31) Dorodnyy, A.; Koepfli, S. M.; Lochbaum, A.; Leuthold, J. Design of CMOS-Compatible Metal–Insulator–Metal Metasurfaces via Extended Equivalent-Circuit Analysis. *Sci Rep* **2023**, *10* (1), 1–12. https://doi.org/10.1038/s41598-020-74849-5.

(32) Han, F.; Liao, S.; Yuan, S.; Wu, R.; Zhao, Y.; Xie, Y.; Qin, F.; Ding, L.; Zhang, L.; Monticone, F.; Chum, C. C.; Deng, J.; Mei, S.; Li, Y.; Teng, J.; Hong, M.; Zhang, S.; Alù, A.; Qiu, C.-W. Hybrid Bilayer Plasmonic Metasurface Efficiently Manipulates Visible Light. *Sci. Adv.* **2016**, *2* (1), e1501168. https://doi.org/10.1126/sciadv.1501168.

(33) Wang, Z.; Bovik, A. C.; Sheikh, H. R.; Simoncelli, E. P. Image Quality Assessment: From Error Visibility to Structural Similarity. *IEEE Trans. on Image Process.* **2004**, *13* (4), 600–612. https://doi.org/10.1109/TIP.2003.819861.

(34) He, K.; Zhang, X.; Ren, S.; Sun, J. Deep Residual Learning for Image Recognition. In *2016 IEEE Conference on Computer Vision and Pattern Recognition (CVPR)*; 2016; pp 770–778. https://doi.org/10.1109/CVPR.2016.90.

(35) Zandehshahvar, M.; Kiarashi, Y.; Zhu, M.; Bao, D.; H Javani, M.; Pourabolghasem, R.; Adibi, A. Metric Learning: Harnessing the Power of Machine Learning in Nanophotonics. *ACS Photonics* **2023**, *10* (4), 900–909. https://doi.org/10.1021/acsphotonics.2c01331.

(36) Abdi, H.; Williams, L. J. Principal Component Analysis. *WIREs Computational Stats* **2010**, *2* (4), 433–459. https://doi.org/10.1002/wics.101.

(37) Roweis, S. T.; Saul, L. K. Nonlinear Dimensionality Reduction by Locally Linear Embedding. *Science* **2000**, *290* (5500), 2323–2326. https://doi.org/10.1126/science.290.5500.2323.

(38) Hinton, G. E.; Roweis, S. Stochastic Neighbor Embedding. In *Advances in Neural Information Processing Systems*; MIT Press, 2002; Vol. 15.

(39) Bank, D.; Koenigstein, N.; Giryes, R. Autoencoders. In *Machine Learning for Data Science Handbook: Data Mining and Knowledge Discovery Handbook*; Rokach, L., Maimon, O., Shmueli, E., Eds.; Springer International Publishing: Cham, 2023; pp 353–374. https://doi.org/10.1007/978-3-031-24628-9_16.

(40) Liu, C.; Maier, S. A.; Li, G. Genetic-Algorithm-Aided Meta-Atom Multiplication for Improved Absorption and Coloration in Nanophotonics. *Photon. Res.* **2020**, *7* (7), 1716–1722. https://doi.org/10.1021/acsphotonics.0c00266.

(41) Thompson, J. R.; Nelson-Quillin, H. D.; Coyle, E. J.; Vernon, J. P.; Harper, E. S.; Mills, M. S. Particle Swarm Optimization of Polymer-Embedded Broadband Metasurface Reflectors. *Opt. Express* **2021**, *29* (26), 43421. https://doi.org/10.1364/OE.444112.

(42) Elsawy, M. M. R.; Lanteri, S.; Duvigneau, R.; Brière, G.; Mohamed, M. S.; Genevet, P. Global Optimization of Metasurface Designs Using Statistical Learning Methods. *Sci Rep* **2017**, *9* (1), 17918. https://doi.org/10.1038/s41598-019-53878-9.

(43) Liu, X.; Wang, P.; Xiao, C.; Fu, L.; Xu, J.; Zhang, D.; Zhou, H.; Fan, T. Compatible Stealth Metasurface for Laser and Infrared with Radiative Thermal Engineering Enabled by Machine Learning. *Adv Funct Materials* **2023**, *33* (11), 2212068. https://doi.org/10.1002/adfm.202212068.

(44) Ma, W.; Xu, Y.; Xiong, B.; Deng, L.; Peng, R.; Wang, M.; Liu, Y. Pushing the Limits of Functionality-Multiplexing Capability in Metasurface Design Based on Statistical Machine Learning. *Advanced Materials* **2022**, *34* (16), 2110022. https://doi.org/10.1002/adma.202110022.

# Anchor-Controlled Generative Adversarial Network for High-Fidelity Electromagnetic and Structurally Diverse Metasurface Design

*Yunhui Zeng[1,2], Hongkun Cao[2*], Xin Jin[1,2*],*

[1]Shenzhen International Graduate School, Tsinghua University, Shenzhen 518055, China

[2]Peng Cheng Laboratory, Shenzhen 518055, China

**\*Email:** jin.xin@sz.tsinghua.edu.cn (Xin Jin), caohk@pcl.ac.cn (Hongkun Cao)

# Section S1. The pseudocode of the training of AcGAN

---

**Algorithm 1**: AcGAN Training Procedure for Metasurface Design

---

**Input:** Training pairs $\{(M_r^i, s_t^i, u_i)\}$, where $M_r^i$ are referenced metasurface, $s_t^i$ are target spectral samples and $u_i$ are control vectors.

**Parameters**: Number of epochs $E$, batch size $B$, learning rate $\eta$, spectral and structural loss weights $\alpha, \beta$, adversarial and mismatch loss weight $\gamma$

1. Initialize parameters $\theta_G$ for generator $G$ and parameters $\theta_D$ for discriminator $D$ with random weights
2. Pre-train AnchorNet to predict spectral properties from metasurface designs;
3. **For** epoch = 1 to $E$ **do**
4.     Shuffle the training data;
5.     Sample minibatch of $m$ noise samples $\{z^{(1)}, \dots, z^{(m)}\}$ from $p_Z(z)$; sample minibatch of $m$ control vectors $\{u^{(1)}, \dots, u^{(m)}\}$ from $p_U(u)$; sample minibatch of m examples $\{(M_r^{(1)}, s_t^{(1)}), \dots, (M_r^{(m)}, s_t^{(m)})\}$ from $p_{\text{data}}(M_r, s_t)$;
6.     **For** each batch $\{(z, u, x)\}$ of size $B$ **do**
        `//Update the discriminator`
7.         **For** $k$ steps:
8.             $M_g \leftarrow G(z, u)$ `//Generate metasurface conditioned on u;`
9.             $D_{\text{real}} \leftarrow D(x, u)$ `//Discriminator output for referenced metasurface;`
10.            $D_{\text{fake}} \leftarrow D(M_g, u)$ `//Discriminator output generated metasurface;`
11.            $D_{\text{mismatch}} \leftarrow D(M_g, u')$ `//Discriminator output for mismatched data;`
12.            $L_D \leftarrow \gamma(L_{adv}^D + L_{\text{mismatch}}) + \alpha L_{\text{spectral}}$ ; `//Calculate the loss for D;`
13.            $\theta_D' \leftarrow \theta_D - \eta \nabla L_D$; `//Update D by ascending its stochastic gradient;`
14.         **End for**
        `//Update the generator`
15.         $M_g \leftarrow G(z, u)$ `//Generate metasurface conditioned on u;`
16.         $D_{\text{fake}} \leftarrow D(M_g, u)$ `//Discriminator output generated metasurface;`
17.         $L_G = \gamma L_{adv}^G + \alpha L_{\text{spectral}} + \beta L_{\text{structural}}$ ; `//Calculate the loss for G;`
20.         $\theta_D' \leftarrow \theta_D - \eta \nabla L_D$; `//Update G by ascending its stochastic gradient;`
21.     **End for**
21. **End for**

**Output:** Trained models D and G.

**Section S2. Literature review on spectral similarity evaluation metrics**

Table S1. Literature review of evaluation matrix of spectral similarity

| Research Work | Year | Journal | Inverse/Forward | matrix |
|---|---|---|---|---|
| So S, Rho J.[1] | 2019 | Nanophotonics | Inverse | MAE |
| Ma W, Cheng F, Xu Y, et al.[2] | 2019 | Advanced Materials | Inverse | MSE |
| Yeung C, Tsai J M, King B, et al.[3] | 2020 | ACS Photonics | Forward | MSE |
| Yeung C, Tsai R, Pham B, et al.[4] | 2021 | Advanced Optical Materials | Inverse | MSE |
| Han X, Fan Z, Liu Z, et al.[5] | 2021 | InfoMat | Inverse | MSE |
| Yeung C, Tsai J M, King B, et al.[6] | 2021 | Nanophotonics | Inverse | MSE |
| Mekki-Berrada F, Ren Z, Huang T, et al.[7] | 2021 | npj Computational Materials | Inverse | MSE |
| Tanriover I, Lee D, Chen W, et al.[8] | 2022 | ACS Photonics | Inverse | MSE |
| Patel S K, Parmar J, Katkar V.[9] | 2022 | Renewable Energy | Inverse | MAE |
| Zhang J, Qian C, Fan Z, et al.[10] | 2022 | Advanced Optical Materials | Inverse | MSE |
| Liu X, Wang P, Xiao C, et al.[11] | 2023 | Advanced Functional Materials | Inverse | MSE |

In existing works, Mean Squared Error (MSE) and Mean Absolute Error (MAE) are the prevalent metrics for assessing spectral similarity, defined as MSE $(s_t, s_g) = \frac{1}{n}\sum_{i=1}^{n}(s_t - s_g)^2$, and · MAE $(s_t, s_g) = \frac{1}{n}\sum_{i=1}^{n}|s_t - s_g|$, respectively. While these metrics are favored in deep learning-based metasurface design due to their straightforward computational properties and robust performance, they often fail to capture critical nuances of metasurface spectral features, such as resonance peaks or specific absorption bands that are crucial yet may constitute only a fraction of the overall spectrum. This oversight can lead to deceptively favorable evaluations, as these minute but essential spectral details are overlooked. To overcome these limitations, our work propose SOC, unlike MSE and MAE, SOC directly quantifies the extent of spectral overlap, offering a more granular and accurate measure of spectral congruence. This metric enhances the assessment of metasurface designs by ensuring that all significant spectral features, especially those critical for specific applications, are accurately matched. The introduction of SOC represents a significant advancement in the field, providing a metric that is not only mathematically sound but also specifically tailored to the unique requirements of metasurface spectral evaluation.

**Reference:**
(1) *Designing nanophotonic structures using conditional deep convolutional generative adversarial networks*. https://www.degruyter.com/document/doi/10.1515/nanoph-2019-0117/html (accessed 2022-12-11).
(2) Ma, W.; Cheng, F.; Xu, Y.; Wen, Q.; Liu, Y. Probabilistic Representation and Inverse Design of Metamaterials Based on a Deep Generative Model with Semi-Supervised Learning Strategy. *Advanced Materials* **2019**, *31* (35), 1901111. https://doi.org/10.1002/adma.201901111.
(3) Yeung, C.; Tsai, J.-M.; King, B.; Kawagoe, Y.; Ho, D.; Knight, M. W.; Raman, A. P. Elucidating the Behavior of Nanophotonic Structures through Explainable Machine Learning Algorithms. *ACS Photonics* **2020**, *7* (8), 2309–2318. https://doi.org/10.1021/acsphotonics.0c01067.
(4) Yeung, C.; Tsai, R.; Pham, B.; King, B.; Kawagoe, Y.; Ho, D.; Liang, J.; Knight, M. W.; Raman, A. P. Global Inverse Design across Multiple Photonic Structure Classes Using Generative Deep Learning. *Advanced Optical Materials* **2021**, *9* (20), 2100548.
(5) Han, X.; Fan, Z.; Liu, Z.; Li, C.; Guo, L. J. Inverse Design of Metasurface Optical Filters Using Deep Neural Network with High Degrees of Freedom. *InfoMat* **2021**, *3* (4), 432–442. https://doi.org/10.1002/inf2.12116.
(6) Yeung, C.; Tsai, J.-M.; King, B.; Pham, B.; Ho, D.; Liang, J.; Knight, M. W.; Raman, A. P. Multiplexed Supercell Metasurface Design and Optimization with Tandem Residual Networks. *Nanophotonics* **2021**, *10* (3), 1133–1143. https://doi.org/10.1515/nanoph-2020-0549.
(7) Mekki-Berrada, F.; Ren, Z.; Huang, T.; Wong, W. K.; Zheng, F.; Xie, J.; Tian, I. P. S.; Jayavelu, S.; Mahfoud, Z.; Bash, D.; Hippalgaonkar, K.; Khan, S.; Buonassisi, T.; Li, Q.; Wang, X. Two-Step Machine Learning Enables Optimized Nanoparticle Synthesis. *npj Comput Mater* **2021**, *7* (1), 55. https://doi.org/10.1038/s41524-021-00520-w.
(8) Tanriover, I.; Lee, D.; Chen, W.; Aydin, K. Deep Generative Modeling and Inverse Design of Manufacturable Free-Form Dielectric Metasurfaces. *ACS Photonics* **2022**. https://doi.org/10.1021/acsphotonics.2c01006.
(9) Patel, S. K.; Parmar, J.; Katkar, V. Graphene-Based Multilayer Metasurface Solar Absorber with Parameter Optimization and Behavior Prediction Using Long Short-Term Memory Model. *Renewable Energy* **2022**, *191*, 47–58. https://doi.org/10.1016/j.renene.2022.04.040.
(10) Zhang, J.; Qian, C.; Fan, Z.; Chen, J.; Li, E.; Jin, J.; Chen, H. Heterogeneous Transfer-Learning-Enabled Diverse Metasurface Design. *Advanced Optical Materials* **2022**, *10* (17), 2200748. https://doi.org/10.1002/adom.202200748.
(11) Liu, X.; Wang, P.; Xiao, C.; Fu, L.; Xu, J.; Zhang, D.; Zhou, H.; Fan, T. Compatible Stealth Metasurface for Laser and Infrared with Radiative Thermal Engineering Enabled by Machine Learning. *Adv Funct Materials* **2023**, *33* (11), 2212068. https://doi.org/10.1002/adfm.202212068.

**Section S3. The detailed hyperparameter setting of AnchorNet**

Table S2. The detailed hyperparameter setting of AnchorNet

| Hyperparameter | Value |
| --- | --- |
| $C \times H \times W$ | $3 \times 64 \times 64$ |
| N | 800 |
| Learning rate | 0.001 |
| Batch size | 64 |
| Maximum training epochs | 500 |
| Loss function | SOC |
| Early stopping patience | 30 |
| Optimizer | Adam |
| Betas | (0.9, 0.999) |
| Epsilon | $1 \times 10^{-8}$ |
| Valid step | 1 |

Training was initially capped at 500 epochs. However, training was stopped early at 312 epochs due to no improvement in the validation set over 30 consecutive epochs, as measured by the SOC) This decision was based on our empirical determination of hyperparameters, including a learning rate of 0.001, a batch size of 64, and the Adam optimizer with beta values of (0.9, 0.999). These settings align with standard practices in deep learning to optimize training efficiency and effectiveness.

## Section S4. Evaluating the impact of spectral similarity evaluation metrics on spectral data dimensionality reduction

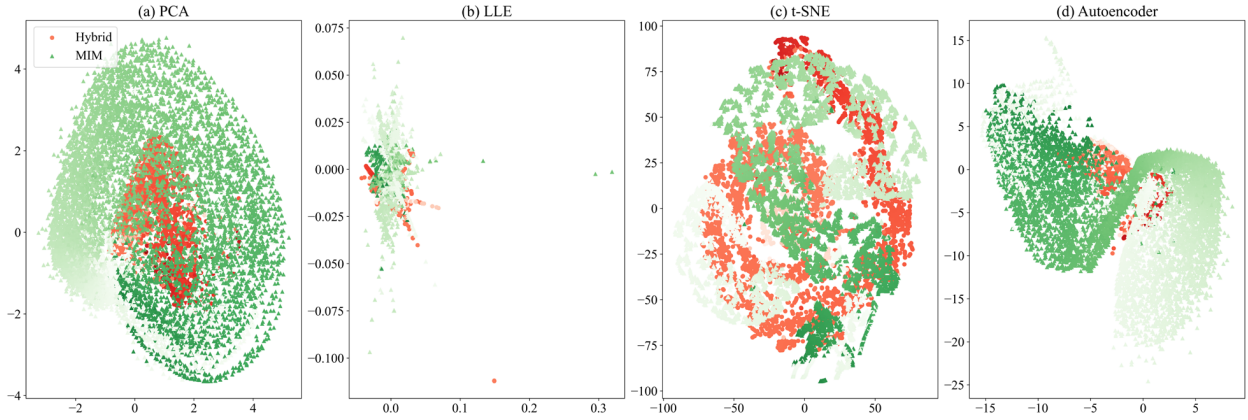

**Figure S1**. Performance of dimensionality reduction techniques using MSE: This figure shows the results of four dimensionality reduction methods (PCA, LLE, t-SNE, and Autoencoder) using MSE as the loss function. (a) PCA, (b) LLE, (c) t-SNE, and (d) Autoencoder depict the embeddings of Hybrid (red circles) and MIM (green triangles) spectral types. These plots highlight MSE's limitations in distinguishing between the two spectral classes across all methods.

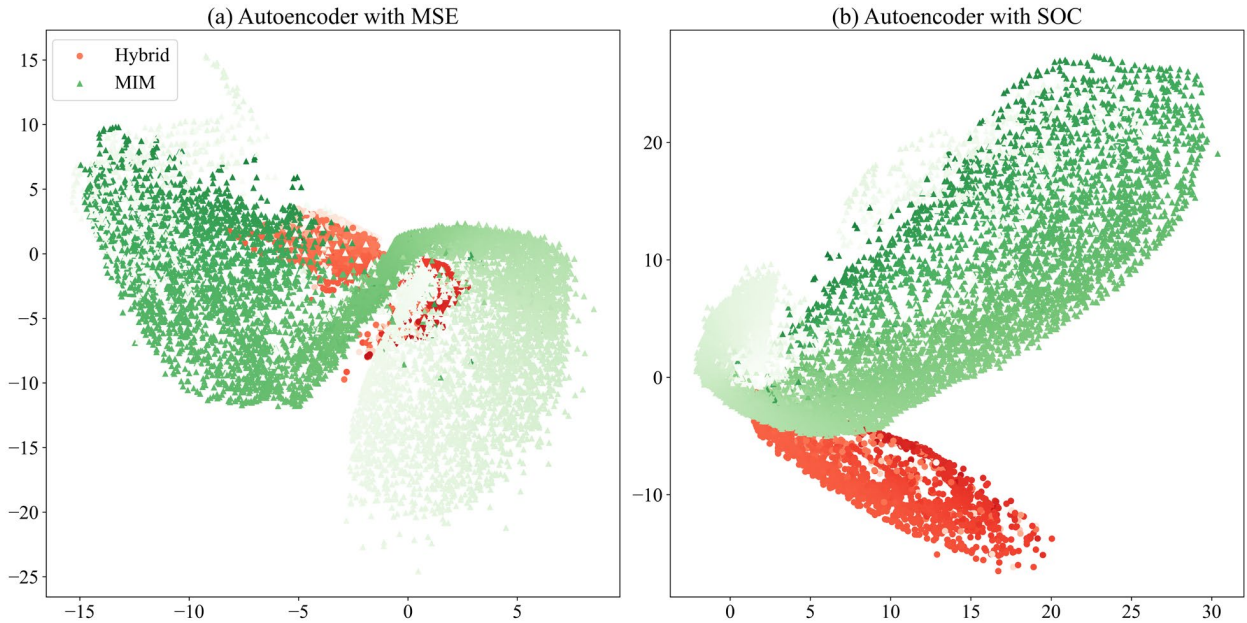

**Figure S2**. Improved spectral data dimensionality reduction with SOC using Autoencoders: (a) and (b) illustrate the effect of replacing MSE with SOC as the loss function in an autoencoder for spectral data dimensionality reduction. (a) shows the results using MSE, with significant overlap between Hybrid (red circles) and MIM (green triangles) spectra. (b) shows the results using SOC, highlighting a clear separation between Hybrid and MIM spectra, confirming SOC's effectiveness in improving spectral class differentiation and preserving intrinsic spectral properties.

In this study, we focus on exploring the effects of different loss functions on the dimensionality reduction and classification of spectral data, with a particular emphasis on the comparative performance between traditional MSE and SOC. Our experiments initially employed four mainstream dimensionality reduction techniques: Principal Component Analysis (PCA), Locally Linear Embedding (LLE), t-Distributed Stochastic Neighbor Embedding (t-SNE), and Autoencoders (AE), using MSE as the loss function. The results demonstrated that regardless of the dimensionality reduction method used, the two spectral classes (Hybrid and MIM) remained indistinguishable with MSE, highlighting its limitations in preserving spectral characteristics.

To further validate SOC's efficacy in dimensionality reduction and classification, we conducted additional experiments with an autoencoder to compare the effects of MSE and SOC as loss functions. As illustrated in Figure S3, when SOC was employed as the loss function, there was a clear separation and formation of distinct clusters for the Hybrid and MIM data in the reduced dimension space. This demonstrated SOC's effectiveness in distinguishing spectral responses and revealing intrinsic data features. SOC also preserved physical characteristics, aiding generalization for resonant responses with slight wavelength variations—key for applications like laser cavity design and photonic crystal optimization. Overall, SOC provides a valuable approach for analyzing spectral data and enhancing optical device performance in nanophotonics and optical design.

**Section S5. The detailed hyperparameter setting of AcGAN**

Table S3. The detailed hyperparameter setting of AcGAN

| Hyperparameter | Value |
|---|---|
| Learning rate | 0.001 |
| Batch size | 256 |
| Training epochs | 1000 |
| Loss function | SOC |
| Optimizer | Adam |
| Betas | (0.9, 0.999) |
| Epsilon | $1 \times 10^{-8}$ |
| Latent vector length | 800 |
| $k$ | 2 |
| $\alpha$, | 0.1 |
| $\beta$ | 0.1 |
| $\gamma$ | 0.9 |

The AcGAN was rigorously trained to optimize metasurface designs. Training was conducted using the Adam optimizer with a learning rate of 0.001 over 1000 epochs and a batch size of 256. The SOC, assessing the precision in replicating target spectral features, served as the core metric for model evaluation and optimization. Initially, the spectral and structural loss weights ($\alpha, \beta$) were set at 0.1, while the adversarial and mismatch loss weight ($\gamma$) was set at 0.9. The loss weights were adjusted incrementally according to the formula:

$$\alpha = \beta = \alpha_0 \times r^t$$
$$\gamma = 1 - \alpha - \beta \tag{S1}$$

where $\alpha_0$ was set to 0.1 and $r$ to 1.002. This dynamic adjustment gradually shifts focus from adversarial robustness to spectral and structural accuracy as training progresses. The model uses a latent vector length of 800 and a clustering mechanism with $k = 2$ to refine the training. The Adam optimizer's beta parameters were set at (0.9, 0.999) to balance fast convergence with training stability, crucial for learning complex metasurface designs.

**Section S6. Diverse near-field electric responses of metasurfaces with similar absorption spectra designed by AcGAN**

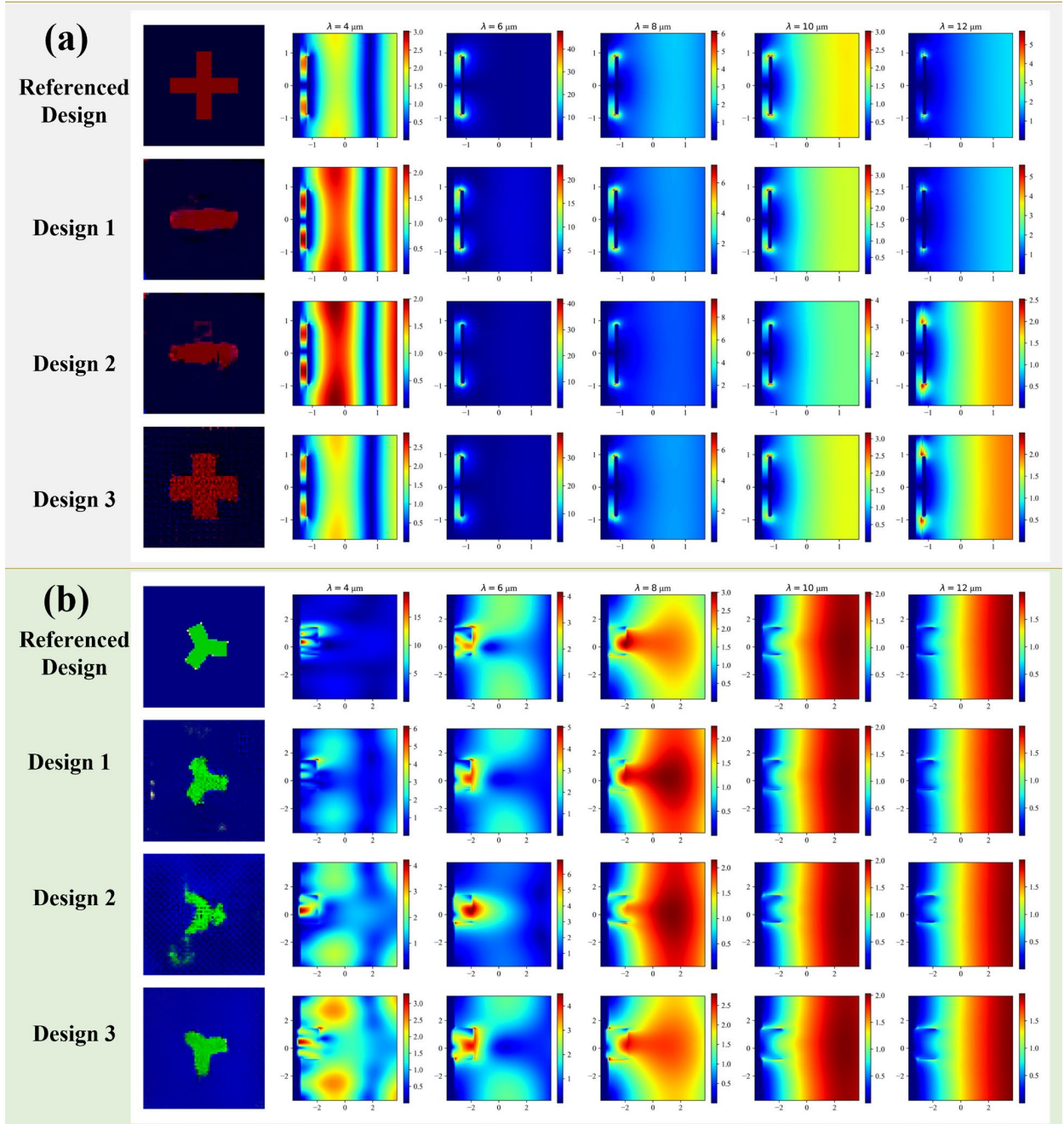

**Figure S3.** Near-field electric responses in the XZ plane for MIM and hybrid metasurface: (a) MIM metasurface: Showcases the near-field electric responses at various wavelengths (4 *μm* to 12 *μm*) for MIM metasurface in **Figure 5** (f). The depth of the colors i reflects the magnitude of the near-field electric field strength. (b) Hybrid metasurface: Presents the near-field electric responses at wavelengths from 4 *μm* to 12 *μm* for hybrid metasurface in **Figure 5** (i).

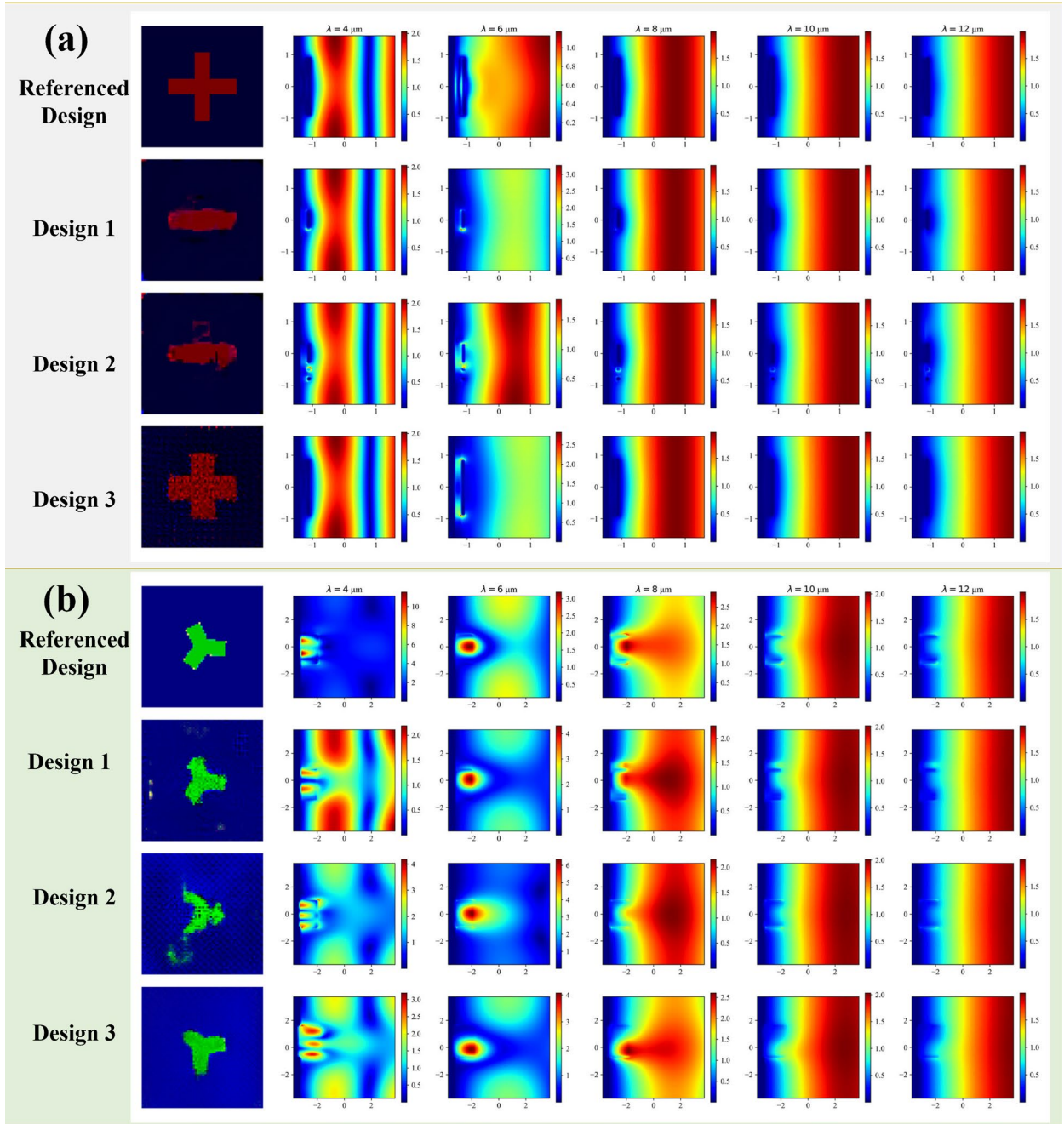

**Figure S4.** Near-field electric responses in the YZ plane for MIM and hybrid metasurface: (a) MIM metasurface: Showcases the near-field electric responses at various wavelengths (4 *μm* to 12 *μm*) for MIM metasurface in **Figure 5** (f). The depth of the colors i reflects the magnitude of the near-field electric field strength. (b) Hybrid metasurface: Presents the near-field electric responses at wavelengths from 4 *µm* to 12 *µm* for hybrid metasurface in **Figure 5** (i).

The primary goal of our experiment is to evaluate the capability of AcGAN in designing metasurfaces that achieve both high electromagnetic fidelity and structural diversity, tailored to diverse electromagnetic applications. Specifically, we aim to demonstrate that AcGAN can generate metasurfaces with similar absorption spectra but varied near-field electric responses when analyzed across different spatial orientations (XY, XZ, and YZ planes). Due to space constraints, we have selected representative examples from the MIM and Hybrid structures, as shown in Figure 5(f) and Figure 5(i), to illustrate the near-field electric responses.

These results reveal the magnitude of the electric fields across different spatial orientations (XY, XZ, and YZ planes), emphasizing regions of high and low EM activity. The images illustrate how the electric field varies across different planes, which is crucial for understanding energy distribution and the metasurface's effectiveness in manipulating EM waves. Each figure includes a color scale that represents the magnitude of the near-field electric response, with warmer colors indicating higher intensity and cooler colors representing lower intensity. The color bar quantifies the range of field intensities, allowing for a precise understanding of the field's strength across different regions of the metasurface. Analysis of these figures shows that, despite having similar absorption spectra, the metasurfaces exhibit distinct near-field patterns across different orientations. This validates the hypothesis that AcGAN can introduce structural variability without compromising EM performance. This finding is significant as it highlights AcGAN's potential in designing advanced optical devices where functional diversity is required alongside specific optical properties. This capability extends the utility of metasurfaces beyond conventional applications, enabling the creation of customized solutions for complex optical systems.

**Section S7. Generation of four arbitrarily-defined spectra**

To assess the AcGAN framework's capacity for designing metasurfaces corresponding to different spectral profiles, we synthesized four types of spectra, each crafted using specific mathematical expressions with defined parameter ranges. These spectra are discretized into 800 points, simulating measurements across a wavelength range from $4\mu m$ to $12\mu m$ :

1. **Gaussian Spectrum Generation**

Gaussian spectra are constructed as follows:

$$s(i) = A\exp\left(-\frac{(i-\mu)^2}{2\sigma^2}\right) \tag{S2}$$

where $i$ indexes the discretized points. The mean $\mu$ varies randomly within $[-2,2]$, the standard deviation $\sigma$ within $[0.5,2]$, and the amplitude $A$ is scaled between $[0.5,1.0]$ to normalize the peak amplitude.

2. **Gate-shaped Spectrum Generation**

Gate-shaped spectra are defined by a constant amplitude over a random interval:

$$s(i) = \begin{cases} A & \text{if } l \leq i \leq u \\ 0 & \text{otherwise} \end{cases} \tag{S3}$$

The boundaries $l$ and $u$ of the active interval are selected such that the width is between 50 and 200 points, and the amplitude A ranges from 0.6 to 1.0 .

3. **Lorentzian Spectrum Generation**

Lorentzian spectra are described by:

$$s(i) = \frac{A\gamma}{\pi((i-x_0)^2 + \gamma^2)} \tag{S4}$$

The parameter $x_0$, the center of the peak, is randomly chosen from $[-5,5]$, $\gamma$, indicating the width, varies between $[0.1,2]$, and the amplitude $A$ is adjusted to fall within $[0.6,1.0]$.

## 4. Fano Resonance Spectrum Generation

Fano resonance profiles are generated using:

$$s(i) = A \frac{(q\gamma + i - x_0)^2}{\gamma^2 + (i - x_0)^2} \tag{S5}$$

where $q$ represents the asymmetry parameter with a range between 3 and 7. Other parameters, $x_0$ and $\gamma$, are set similarly to the Lorentzian spectrum, with amplitude $A$ being normalized similarly.

These synthesized spectra, mapped across 800 data points, serve as test cases for evaluating AcGAN's design capabilities under varied spectral conditions. This setup mimics realistic scenarios by capturing diverse spectral behaviors within the specified wavelength range, ensuring a robust evaluation of the framework's performance in generating metasurface designs tailored to specific electromagnetic properties.

**Section S8. The performance of AcGAN of arbitrarily-defined spectrum**

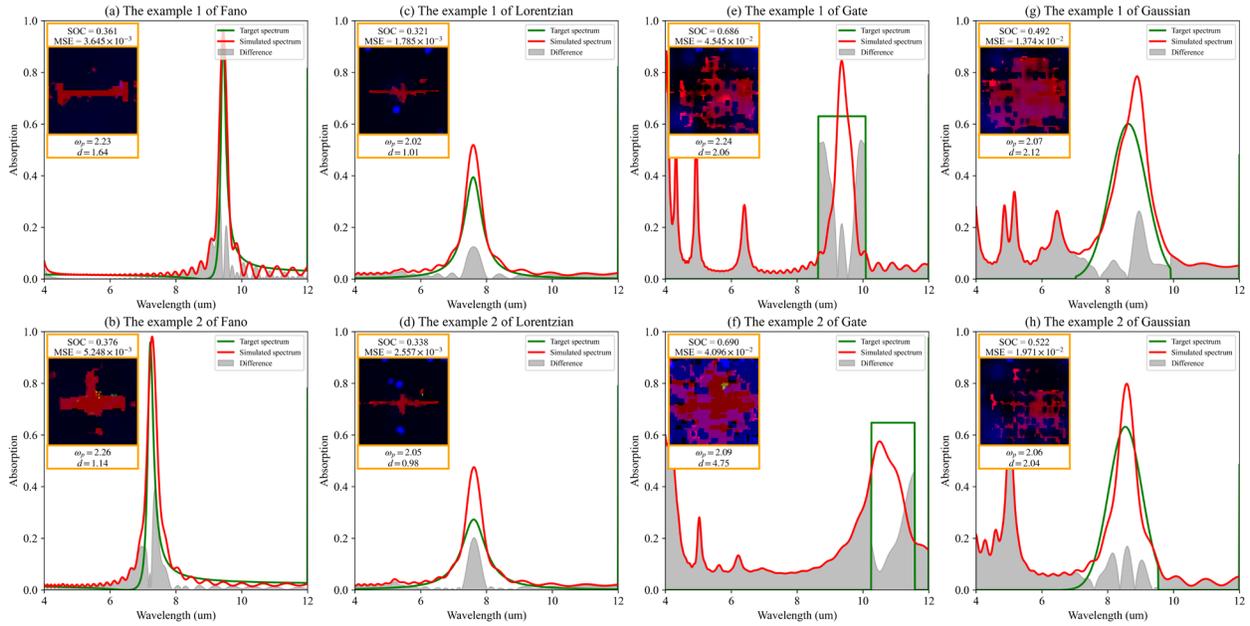

**Figure S5.** AcGAN response to to different spectral types for which there are no corresponding structures in the dataset. For each type, two examples are shown to demonstrate AcGAN's predictive accuracy compared to the target spectra. The simulated spectra (red curves) are compared with the target spectra (green curves) to highlight the model's precision. Each subplot includes SOC and MSE metrics to quantify the discrepancy between the simulated and target spectra. Gray shaded areas indicate significant spectral mismatches. Insets within each panel display the corresponding metasurface structures, illustrating the correlation between spectral properties and physical configurations.

**Section S9. Correlation between weighted distance and SOC for various spectral types**

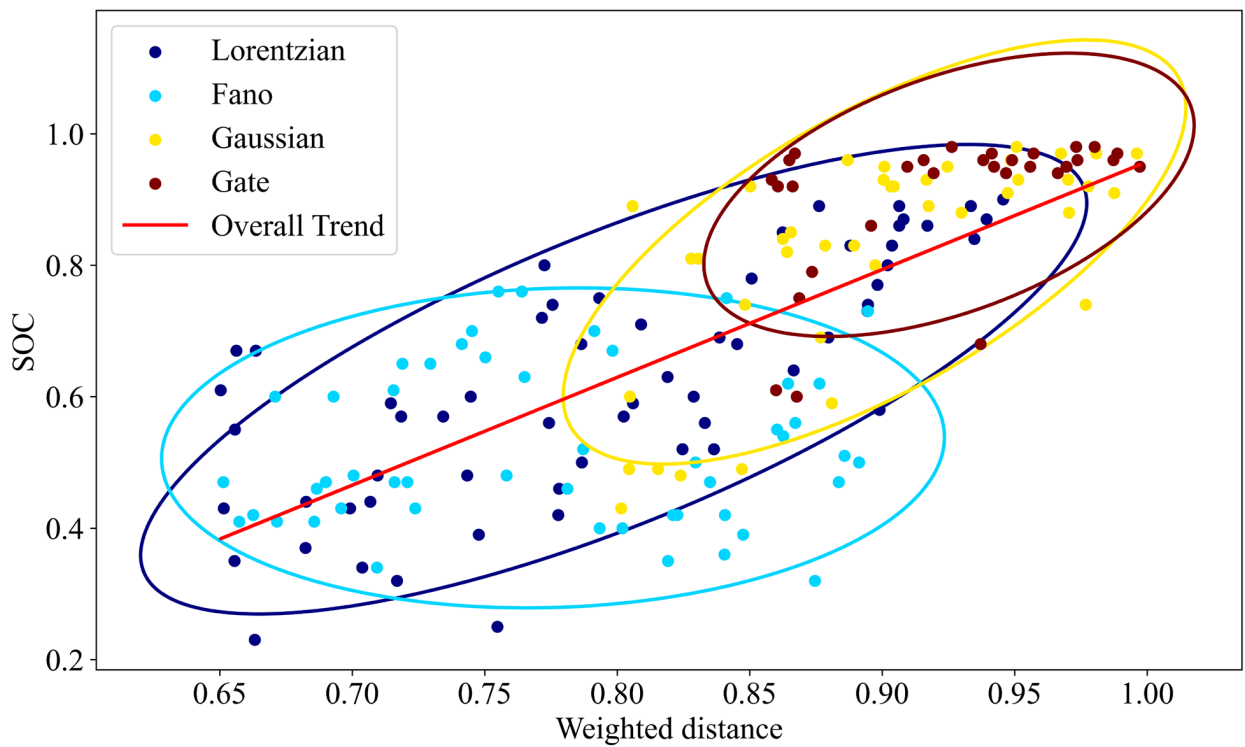

**Figure S6.** Correlation between weighted distance and SOC for various spectral types. The scatter plot shows the relationship between the weighted distance from dataset cluster centers and the SOC for generated spectra. Points represent Gaussian (blue), Gate (yellow), Lorentzian (black), and Fano (red) types. Ellipses represent the data distribution for each type. The overall trend (red line) indicates that closer proximity to training data (lower weighted distance) generally corresponds to higher SOC, indicating worse model performance.

**Section S10. Ablation study**

Table S4. Ablation study comparing AcGAN performances with and without cluster controller

| Turn on the Cluster Controller | Average MSE | Average SOC |
|---|---|---|
| False | $9.770 \times 10^{-3}$ | 0.434 |
| **True** | $\mathbf{1.120 \times 10^{-3}}$ | **0.139** |

The inclusion of the cluster controller significantly enhances model accuracy. With the cluster controller activated, the average MSE decreased from $9.770 \times 10^{-3}$ to $1.120 \times 10^{-3}$, and the SOC improved from 0.434 to 0.139 (Table S4). This stark improvement underscores the controller's ability to refine the spectral dataset's representation, which aids in generating metasurfaces that more closely match the desired spectral characteristics. It suggests that the cluster controller effectively captures pivotal spectral features that are essential for the network to generate high-fidelity designs.

Table S5. Ablation study comparing AcGAN performances with and without AnchorNet integration in generator and discriminator

| Turn on the AnchorNet on the generator | Turn on the AnchorNet on the discriminator | Average MSE | Average SOC |
|---|---|---|---|
| False | False | $2.674 \times 10^{-3}$ | 0.228 |
| True | False | $2.742 \times 10^{-3}$ | 0.224 |
| False | True | $2.512 \times 10^{-3}$ | 0.220 |
| **True** | **True** | $\mathbf{1.120 \times 10^{-3}}$ | **0.139** |

Incorporating AnchorNet within both the generator and the discriminator significantly optimizes the network's performance. The configuration where AnchorNet is active in both the generator and discriminator simultaneously yielded the best results, with an MSE of $1.120 \times 10^{-3}$ and an SOC of 0.139 (Table S5). This dual deployment enables a more cohesive and accurate evaluation of generated metasurfaces against the target spectra, demonstrating that synchronized feedback between these components is crucial for enhancing the fidelity of generated designs.

**Table S6.** Impact of k-value on AcGAN performance

| $k$ value | Average MSE | Average SOC |
|---|---|---|
| 1 | $2.613 \times 10^{-3}$ | 0.207 |
| 2 | $\mathbf{1.120 \times 10^{-3}}$ | **0.139** |
| 4 | $3.522 \times 10^{-3}$ | 0.221 |
| 8 | $9.698 \times 10^{-3}$ | 0.415 |

The balance between discriminator and generator training frequencies, controlled by the k-value, plays a crucial role in stabilizing the adversarial training process. We varied the k-value to adjust how many times the generator updates for each discriminator update. Our findings indicated that a k-value of 2 provided the best performance, minimizing MSE to $1.120 \times 10^{-3}$ and SOC to 0.139, suggesting that too frequent generator updates relative to the discriminator might lead to inefficiencies, potentially due to overfitting or unstable adversarial dynamics.

**Table S7.** Effect of initial spectral loss weight $\alpha$ on AcGAN performance

| Initial spectral loss weights $\alpha$ | Average MSE | Average SOC |
|---|---|---|
| 0.1 | $\mathbf{1.120 \times 10^{-3}}$ | **0.139** |
| 0.2 | $3.003 \times 10^{-3}$ | 0.271 |
| 0.3 | $3.451 \times 10^{-3}$ | 0.309 |
| 0.4 | $3.719 \times 10^{-3}$ | 0.333 |
| 0.5 | $5.405 \times 10^{-3}$ | 0.384 |
| 0.6 | $8.830 \times 10^{-3}$ | 0.466 |
| 0.7 | $1.075 \times 10^{-2}$ | 0.486 |
| 0.8 | $1.023 \times 10^{-2}$ | 0.488 |
| 0.9 | $9.705 \times 10^{-3}$ | 0.475 |
| 1.0 | $1.457 \times 10^{-2}$ | 0.550 |
| 2.0 | $1.297 \times 10^{-2}$ | 0.530 |

The weight of the spectral loss, managed by α, directly impacts how the feedback from the AnchorNet shapes the generator's outputs. Surprisingly, an α of 0.1 yielded the most favorable outcomes, with the lowest MSE and highest SOC, indicating a well-balanced adversarial and spectral loss contribution. This suggests that while spectral alignment is critical, its dominance in the loss function must be carefully calibrated to avoid overshadowing the adversarial component necessary for diverse and generalizable metasurface design.

Table S8. AcGAN performance across various latent sizes

| Latent size | Average MSE | Average SOC |
|---|---|---|
| 100 | $3.116 \times 10^{-3}$ | 0.238 |
| 200 | $3.388 \times 10^{-3}$ | 0.280 |
| 400 | $3.153 \times 10^{-3}$ | 0.236 |
| 800 | $\mathbf{1.120 \times 10^{-3}}$ | **0.139** |
| 1600 | $1.403 \times 10^{-2}$ | 0.529 |

Exploring different sizes of the latent space, we assessed how the dimensionality influences the model's ability to encapsulate and generate diverse metasurface designs. A latent size of 800 was found to be optimal, striking a balance between complexity and manageability, and significantly outperforming smaller and larger sizes in both MSE and SOC. This highlights the importance of an appropriately sized latent space in capturing the necessary variability in metasurface designs without introducing excessive noise or complexity.

Table S9. AcGAN performance variation with different batch sizes

| Batch size | Average MSE | Average SOC |
|---|---|---|
| 64 | $3.102 \times 10^{-3}$ | 0.223 |
| 128 | $3.318 \times 10^{-3}$ | 0.234 |
| 256 | $\mathbf{1.120 \times 10^{-3}}$ | **0.139** |
| 512 | $2.730 \times 10^{-3}$ | 0.213 |

Our analysis extended to the effect of batch size on model performance. Smaller batch sizes generally led to higher MSE and lower SOC, underscoring the benefits of larger batch sizes in stabilizing training updates and gradient estimations. A batch size of 256 demonstrated the best performance, optimizing both MSE and SOC, facilitating more stable and reliable training cycles.